\documentclass[11pt]{article}
\usepackage[preprint]{acl}

\usepackage{times}
\usepackage[T1]{fontenc}
\usepackage[utf8]{inputenc}
\usepackage{microtype}
\usepackage{inconsolata}

\usepackage{url}
\usepackage{graphicx}

\usepackage{tikz}
\usetikzlibrary{arrows.meta,positioning,calc}
\usepackage{amsmath}

\usepackage{threeparttable}
\usepackage{multirow}
\usepackage{amssymb}
\usepackage{tabularx}
\usepackage{booktabs}
\usepackage{xcolor}

\newcommand{\yes}{\checkmark}
\newcommand{\pmark}{$\circ$}
\newcommand{\no}{$\times$}

\title{Security of Deep Research Agents: Attack Surface, Defenses, Benchmarks, and Agenda Seeding against Research Planning}

\author{
 \textbf{Yang Luo},
 \textbf{Guangpu Chen},
 \textbf{Yichen Liu},
 \textbf{Zhaoyu Chen},
\\
 \textbf{Yifeng Gao},
 \textbf{Zhiyang Ding},
 \textbf{Qingni Shen},
 \textbf{Zhonghai Wu}\thanks{Corresponding author: Zhonghai Wu.}
\\
 Peking University
\\
 \small{\texttt{\{luoyang,qingnishen,wuzh\}@pku.edu.cn}}
\\
 \small{\texttt{\{chenguangpu,liuyichen,2001210212,yfgao25,dingzhiyang26\}@stu.pku.edu.cn}}
}

\begin{document}

\maketitle

\begin{abstract}
Deep research and agentic search agents autonomously retrieve content from the open web and generate reports with citations, and their outputs are often taken directly as research conclusions. This paper systematizes security research in this area along four stages, namely retrieval, planning, evidence integration, and report generation. We present a unified threat model, a taxonomy of attacks and defenses, and a review of evaluation benchmarks, and we align the area dimension by dimension with retrieval-augmented generation (RAG) poisoning, prompt injection, and generative engine optimization (GEO). Coding 34 primary studies shows that only 26.5\% of them involve the planning stage, and none performs an end-to-end poisoning evaluation on commercial products. To address this gap, we propose the agenda seeding attack, in which the attacker writes no instructions and only introduces a fictitious evaluation metric. Across 3{,}520 sandboxed research runs under the same budget, GEO and prompt injection are nearly ineffective, whereas agenda seeding makes the attacker's target the top pick in 16\% to 28\% of reports. Injection detection and query anchoring fail against it, verification prompting hurts utility, and our proposed provenance tagging preserves utility but can be bypassed by multi-site bait. Finally, we propose eight research directions.
\end{abstract}

\section{Introduction}
\label{sec:Introduction}

Starting from WebGPT \citep{nakano2021webgpt} and ReAct \citep{yao2023react}, large language models (LLMs) have been allowed to call search engines and read web pages during reasoning. Subsequently, Search-o1 \citep{li2025searcho1} embedded retrieval into the thinking process of reasoning models, and works such as Search-R1 \citep{jin2025searchr1}, DeepResearcher \citep{zheng2025deepresearcher}, WebThinker \citep{li2025webthinker}, WebSailor \citep{li2025websailor}, and WebDancer \citep{wu2025webdancer} further used reinforcement learning to train models to decide autonomously when to search, what to search for, and how deeply to read. On the product side, OpenAI Deep Research \citep{openai2025deepresearch}, Gemini Deep Research \citep{google2024geminidr}, and Perplexity Deep Research \citep{perplexity2025dr} can visit hundreds of web pages within minutes to tens of minutes and output structured research reports with citations. The academic community has also open-sourced frameworks such as STORM \citep{shao2024storm}, Co-STORM \citep{jiang2024costorm}, OmniThink \citep{xi2025omnithink}, and MindSearch \citep{chen2025mindsearch}. We refer to such systems collectively as deep research and agentic search agents, abbreviated as DR agents in what follows.

The value of a DR agent lies in reading, on the user's behalf, a large number of web pages that the user would never read, which also means that anyone who can place content into its retrieval results can influence a report that the user treats as a research conclusion. Several primary studies since 2025 have made this risk concrete. Agent-Fence \citep{puppala2026agentfence} organizes 14 attack classes by trust boundary and finds, on 8 deep agent archetypes, that wallet draining, authorization confusion, retrieval poisoning, and planning manipulation pose the highest risks. WARP \citep{zhang2026warp} rewrites only about 13 words on a single frequently retrieved user-generated content (UGC) page, which causes the page to be retrieved in 57\% to 76\% of research runs and cited in 38\% to 51\% of reports. HAE-GEO \citep{bi2026haegeo} constructs hierarchical web evidence poisoning with three progressive levels and finds that fabricated mutual corroboration weakens the agent's ability to recognize suspicious evidence. A jailbreak study targeting deep research \citep{chen2025deeperharm} shows that harmful requests that would be refused on their own can, after plan injection or reframing as academic research, yield harmful reports that are more structured and more actionable than those produced in ordinary chat.

However, these works are scattered across security, natural language processing, information retrieval, and machine learning venues and preprints, and they use different terminology, threat models, and metrics. Existing surveys on search agents and deep research \citep{xi2025deepsearch,zhang2025drsurvey,huang2025drsystematic,xu2025comprehensive,li2025aisearch} systematically review paradigms, optimization, and evaluation, but they focus only on capability. Existing agent security surveys \citep{deng2025threat,he2024security,he2025emerged,gan2024navigating,yu2025trustworthy,kim2026landscape,chu2026systematic,ling2026toward} organize threats by model, memory, tools, and environment and treat retrieval as one of many untrusted input channels. They characterize neither the open-web evidence ecosystem (publishers, UGC platforms, ranking, and search services) nor the risks that arise when reports are taken directly as conclusions. In adjacent areas, retrieval-augmented generation (RAG) poisoning \citep{zou2025poisonedrag,zhong2023poisoning}, prompt injection \citep{greshake2023more,liu2024formalizing}, and generative engine optimization (GEO) \citep{aggarwal2024geo,nestaas2025adversarial} have each accumulated many results, but whether their threat models transfer to multi-round, open research pipelines in which queries are autonomously rewritten has not been systematically answered.

This paper systematizes security research on DR agents and conducts a controlled experiment targeting a gap identified during the systematization. The paper is organized around four stages, namely retrieval, planning, evidence integration, and report generation, and answers five research questions (RQs). RQ1: What are the attack surface and threat model of DR agents? RQ2: How can existing attacks and defenses be categorized by stage, and what assumptions does each rely on? RQ3: What do existing benchmarks measure and what do they miss, and how do they relate to RAG poisoning, prompt injection, and GEO? RQ4: What research gaps does the overall distribution of the literature reveal? RQ5: At the planning stage, which literature coding shows to be the most neglected, can an attacker who uses no instructions and only changes the comparison criteria change the conclusions of a DR agent, and can existing defenses at each layer stop it? Our contributions are as follows:

\begin{itemize}
\item We present a four-stage system model and a unified threat model for DR agents, including five trust boundaries, six attacker capability levels (C0 to C5, with the supply-chain attacker C6 listed separately as a gap), and six classes of security goals, and we explain how its threat assumptions differ from those of RAG, Web agents, and ordinary search agents.
\item We build a stage-organized attack taxonomy (4 stages, 19 attack classes) and a defense taxonomy (5 layers), give for each defense class the assumptions it relies on and its known failure conditions, and review the coverage and metrics of 9 security benchmarks.
\item We align DR agent security dimension by dimension with RAG poisoning, prompt injection, GEO, and classical search engine poisoning, and identify which conclusions transfer and which assumptions break down in research pipelines, in particular that instruction-free data-level manipulation bypasses defenses centered on instruction detection.
\item We code 34 primary studies along six dimensions and quantify the skew of the literature in stages, attacker capabilities, attack goals, evaluation environments, and defense evaluation. On this basis, we propose eight research gaps and future directions and explain how this paper differs from existing agent security surveys.
\item We propose the agenda seeding attack and conduct 3{,}520 paired research runs in a reproducible offline sandbox. Under the same budget, it is significantly stronger than GEO and prompt injection, injection detection and root-query anchoring fail against it, and verification prompting can lower its success rate only at a clear cost in utility. We also propose provenance tagging as a lightweight defense that does not degrade research quality, and we use an adaptive attack to expose its limitations.
\end{itemize}

The rest of the paper is organized as follows: Section \ref{sec:Background} introduces the pipeline and system landscape of DR agents, Section \ref{sec:Methodology} describes literature collection and coding, Section \ref{sec:Threat Model} presents the threat model, Sections \ref{sec:Attacks} and \ref{sec:Defenses} present the attack and defense taxonomies, Section \ref{sec:Benchmarks} reviews evaluation benchmarks, Section \ref{sec:Adjacent} establishes connections to adjacent areas, Section \ref{sec:Coding} reports literature coding statistics, Section \ref{sec:Study} reports the agenda seeding experiment, Section \ref{sec:Gaps} proposes research gaps and directions, Section \ref{sec:Surveys} explains the differences from existing surveys, and Section \ref{sec:Conclusion} concludes.

\section{Background}
\label{sec:Background}

\subsection{The Research Pipeline}

\begin{figure*}[tbp]
\centering
\includegraphics[width=0.8\linewidth]{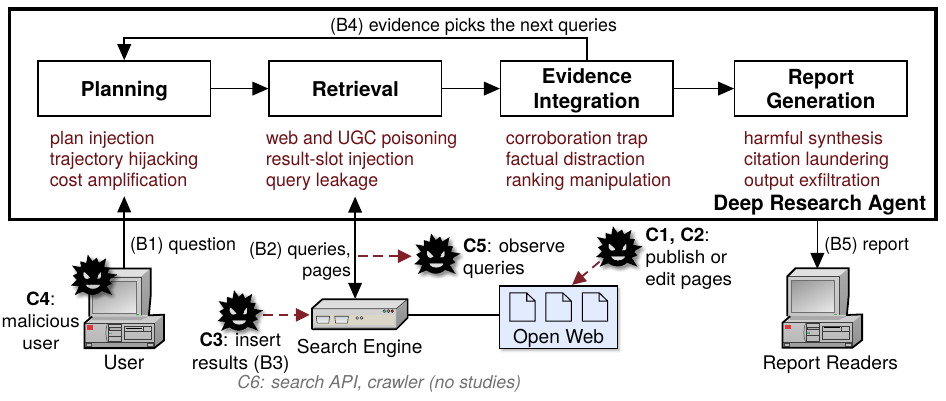}
\caption{The four-stage pipeline of a deep research agent, its trust boundaries (B1 to B5), representative attacks at each stage (red), and where each attacker level (C1 to C6) acts. Dashed red arrows are attacker actions.}
\label{fig_1_pipeline}
\end{figure*}

A DR agent receives an open-ended research question and, after multiple rounds of planning, retrieval, and integration, generates a report. Implementations vary widely across systems, but all of them can be abstracted into the four stages in Figure \ref{fig_1_pipeline}. An existing capability survey \citep{zhang2025drsurvey} adopts a division into planning, question generation, web exploration, and report generation. We merge question generation into planning and split web exploration into retrieval and evidence integration, because from a security perspective, what content is fetched and what content is believed are two different kinds of failure.

\textbf{Retrieval.} The agent issues queries to a search engine or search API, obtains the result list, and then selectively fetches web pages, PDFs, or UGC. Retrieval can be one-shot (as in ordinary search-augmented question answering \citep{vu2024freshllms}) or repeated tens to hundreds of times within a single research session. The retrieval stage spans the widest trust boundary between the agent and the outside world: search ranking is determined by third parties, web content is determined by any publisher, and fetched content enters the model context directly.

\textbf{Planning.} The idea of interleaving retrieval and reasoning can be traced back to Self-Ask \citep{press2023measuring}, IRCoT \citep{trivedi2023interleaving}, and Toolformer \citep{schick2023toolformer}. In a DR agent, the planner decomposes the research question into sub-questions, generates queries for each sub-question, and decides the next step based on what has been read: following up, switching perspectives, digging into a particular source, or stopping. STORM \citep{shao2024storm} organizes research through multi-perspective questioning, MindSearch \citep{chen2025mindsearch} organizes sub-questions as a directed graph, and ``Towards AI Search Paradigm'' \citep{li2025aisearch} further splits planning, execution, and writing across multiple collaborating agents. Search agents trained with reinforcement learning \citep{jin2025searchr1,zheng2025deepresearcher,li2025websailor} make planning implicit in the reasoning trajectory. What makes planning special is that it depends on data: subsequent queries are determined by previously read content, so retrieved content can in turn control the direction of research.

\textbf{Evidence integration.} The agent filters and deduplicates the retrieved evidence, assesses its credibility, resolves conflicts, and extracts key points. This stage determines which claims enter the conclusions. Most existing systems leave credibility judgments to the model, relying on its implicit judgments of source authority, consistency, and timeliness \citep{wang2025astuterag}. On the open web, evidence often conflicts, and agreement among multiple sources does not necessarily imply truth.

\textbf{Report generation.} The agent writes the integrated evidence into a long, sectioned report with citations. Citations make reports appear verifiable, but prior work has found that only about half of the sentences in generative search engine responses are fully supported by their citations \citep{liu2023verifiability}, and generation with citations remains an open problem \citep{gao2023alce}. The report is used directly by users, downstream agents, or decision processes and is the only outlet through which a DR agent affects the outside world.

\subsection{System Landscape}

Table \ref{table_systems} compares DR agents with three classes of adjacent systems. Compared with RAG \citep{lewis2020rag,gao2023ragsurvey}, the corpus of a DR agent is the open web rather than a curated knowledge base, queries are generated by the agent rather than given directly by the user, and retrieval is multi-round and adaptive. Compared with Web agents aimed at completing web operations, such as those evaluated in WebArena \citep{zhou2024webarena} and Mind2Web \citep{deng2023mind2web}, DR agents mostly access the web read-only, and their harms come mainly from report content rather than from side-effecting actions on web pages, although recent products have connected research with execution (sending emails, filling in forms, writing code). Compared with search-augmented question answering (such as Search-o1 \citep{li2025searcho1} and Search-R1 \citep{jin2025searchr1}), DR agents output long reports, and a polluted sub-conclusion may be wrapped among dozens of correct citations, making it harder for readers to notice. For capability evaluation, GAIA \citep{mialon2024gaia}, BrowseComp \citep{wei2025browsecomp}, and DeepResearch Bench \citep{du2025deepresearchbench} measure research capability, but none of them includes an adversary.

\begin{table*}[!tbp]
\centering
\footnotesize
\setlength{\tabcolsep}{3pt}
\caption{Retrieval-Centric LLM Systems and Their Security-Relevant Properties}
\begin{tabularx}{\textwidth}{@{}>{\raggedright\arraybackslash}p{2.6cm}>{\raggedright\arraybackslash}X>{\raggedright\arraybackslash}X>{\raggedright\arraybackslash}X>{\raggedright\arraybackslash}X@{}}
\hline\noalign{\smallskip}
Property & RAG & Web agent & Agentic search & Deep research \\
\noalign{\smallskip}
\hline
\noalign{\smallskip}
Corpus & Curated or private corpus & Live websites under interaction & Open web via search API & Open web, UGC, PDFs, and private documents \\
Retrieval rounds & One per question & Driven by actions & Several, adaptive & Tens to hundreds, adaptive and recursive \\
Query author & User question & Agent & Agent & Planner and sub-agents \\
Output & Short answer & Actions with side effects & Short answer with citations & Long structured report with citations \\
Dominant harm & Wrong answer & Unsafe action & Wrong answer & Misleading report trusted as a research conclusion \\
Examples & RAG \citep{lewis2020rag} & WebArena \citep{zhou2024webarena} & Search-R1 \citep{jin2025searchr1} & STORM \citep{shao2024storm}, OpenAI DR \citep{openai2025deepresearch} \\
\hline
\end{tabularx}
\label{table_systems}
\end{table*}

\section{Systematization Methodology}
\label{sec:Methodology}

\textbf{Scope.} The security research included in this paper targets LLM-driven systems that can autonomously initiate web retrieval and generate answers or reports based on the retrieval results, and covers attacks, defenses, benchmarks, and measurement studies. Pure capability studies, studies targeting only RAG over closed knowledge bases, and studies targeting only Web agents that perform web operations are not counted in the core corpus, but they are included as adjacent work in the alignment analysis of Section \ref{sec:Adjacent}. Studies on manipulating generative engines (LLM ranking and recommendation systems with retrieval) are counted in the core corpus, because they attack the evidence integration and generation stages of DR agents.

\textbf{Sources and search.} We searched the four security conferences USENIX Security, \allowbreak IEEE S\&P, \allowbreak ACM CCS, and \allowbreak NDSS, other venues including ACL, \allowbreak EMNLP, \allowbreak NAACL, \allowbreak NeurIPS, \allowbreak ICLR, \allowbreak ICML, \allowbreak COLM, and \allowbreak KDD, and the arXiv categories cs.CR, \allowbreak cs.CL, \allowbreak cs.IR, and \allowbreak cs.AI, covering January 2023 to September 2026. Keyword combinations paired deep research, \allowbreak search agent, \allowbreak generative engine, and \allowbreak web agent with poisoning, \allowbreak injection, \allowbreak jailbreak, \allowbreak manipulation, \allowbreak leakage, and \allowbreak safety, and we then performed forward and backward snowballing on the matched papers. Since many results in this area are still published as preprints, we included arXiv papers.

\textbf{Inclusion and exclusion.} The inclusion criteria are: (i) the subject of study falls within the above scope, (ii) the work contains an explicit threat model or security goal, and (iii) the work reports empirical results (attack effectiveness, defense effectiveness, or measurement data). We exclude opinion pieces without empirical results, vendor blogs, and news reports. This yields 34 primary studies, plus 88 adjacent works on attacks and defenses, capability systems, and surveys, which are used for the alignment analysis in Section \ref{sec:Adjacent} and for background.

\textbf{Coding scheme.} We code six dimensions for each primary study. (1) Type: attack (A), defense (D), benchmark or evaluation framework (B), or measurement study (M), according to the main contribution. (2) Stages involved: retrieval (R), planning (P), evidence integration (I), and report generation (G), multi-label, determined by where the attacker exerts influence or where the defense acts. (3) Attacker capability: C0 to C5, defined in Section \ref{sec:Threat Model}. (4) Attack goal: integrity of evidence and conclusions (Int), targeted promotion and ranking manipulation (Pro), harmful content generation (Harm), task or tool hijacking (Hij), confidentiality (Conf), and availability and cost (Avail), multi-label. (5) Target system: deep research framework or product (DR), search agent (SA), generative engine (GE), or retrieval-augmented pipeline (RAG). (6) Evaluation environment: the real web (Live), a sandbox that inserts controlled content into real search results (Sandbox), or an offline or simulated corpus (Sim), together with whether defenses were evaluated.

\section{System and Threat Model}
\label{sec:Threat Model}

\subsection{Trust Boundaries}

Figure \ref{fig_1_pipeline} marks the five trust boundaries of a DR agent. B1 is the boundary between the user and the agent, and the user's research request may itself be malicious. B2 is the boundary between the agent and search engines and web pages. Neither web content nor ranking is under the agent's control, and this is the widest and most studied boundary. B3 is the boundary where search result slots and external observers sit: ad slots, compromised search proxies, or third-party search APIs can insert or replace results, and search service providers and network observers can see all queries the agent issues. B4 is the feedback boundary between evidence integration and planning, namely the loop in which content already read determines subsequent queries and subtasks, and in multi-agent systems it also includes delegation and shared memory among sub-agents \citep{gadgil2026badmemory,chen2024agentpoison}. B5 is the boundary between the report and its consumers, where the report is used as a conclusion by humans or downstream agents.

\subsection{Security Goals}

We group the security goals of DR agents into six classes. \emph{Integrity of evidence and conclusions}: factual conclusions in the report should be supported by credible evidence and should not be swayed by claims planted by attackers. \emph{Fairness of ranking and recommendation}: in comparison and recommendation research, the appearance and ranking of entities should not be determined by adversarial optimization by content owners. This is a special case of integrity in commercial settings, listed separately because of its distinctive incentive structure. \emph{Safety}: the agent should not turn harmful requests into actionable harmful reports. \emph{Control-flow integrity}: the agent's planning and tool calls should follow user intent and should not be hijacked by retrieved content. \emph{Confidentiality}: the user's research intent, local private documents, and session data should not leak through queries, requests, or reports. \emph{Availability and cost}: research should not be induced into unbounded loops or into consuming computation and API costs far beyond what the task requires.

\subsection{Attacker Capabilities}

Table \ref{table_attacker} lists attacker levels from weakest to strongest, and Figure \ref{fig_1_pipeline} marks where each of them acts. C1 is the most common assumption: the attacker can publish or control their own web pages but does not know the user's specific queries and cannot control search ranking. C2 can edit existing, frequently retrieved third-party pages, such as forum comments, Q\&A answers, or encyclopedia entries. The key observation of WARP \citep{zhang2026warp} is that such pages already enjoy a ranking advantage, so the attacker does not need to compete with the search engine. C3 can insert entries directly into search results, corresponding to ad slots, compromised search proxies, or malicious search tools. C4 is a malicious user who submits requests directly to the agent. C5 is a passive observer who can see the agent's outbound queries or network traffic. C0 denotes natural failures without an active attacker, such as reference fabrication \citep{rao2026reference}, which causes the same harm as attacks and shares defenses with them, so we include it as well. There is also a stronger class of attacker, C6, namely supply-chain attackers who control search APIs, crawling services, tool servers, or sub-agents. No study in the core corpus addresses it, and we list it as a gap in Section \ref{sec:Gaps}.

\begin{table}[!tbp]
\centering
\footnotesize
\setlength{\tabcolsep}{2.5pt}
\caption{Attacker Capability Levels}
\begin{threeparttable}
\begin{tabularx}{\columnwidth}{@{}l>{\raggedright\arraybackslash}p{2.1cm}>{\raggedright\arraybackslash}X@{}}
\hline\noalign{\smallskip}
Level & Capability & Instances and representative works \\
\noalign{\smallskip}
\hline
\noalign{\smallskip}
C0 & No adversary & Fabricated or mismatched references \citep{rao2026reference} \\
C1 & Publish own pages & Search-optimized and GEO pages, content farms \citep{aggarwal2024geo,bi2026haegeo,zhang2026lazygrounding} \\
C2 & Edit existing third-party pages & UGC comments, wiki or review edits \citep{zhang2026warp,caville2026siren} \\
C3 & Control result slots & Injected search results, ads, malicious search tools \citep{dong2025safesearch,li2026breadcrumbing} \\
C4 & Submit queries & Jailbreak and intent hijacking \citep{chen2025deeperharm,zhan2026safesearch} \\
C5 & Observe queries or traffic & Search providers, network observers \citep{gurung2026mosaicleaks,jeong2025network} \\
C6 & Control supply chain & Search APIs, crawlers, tool servers, sub-agents (no work) \\
\hline
\end{tabularx}
\end{threeparttable}
\label{table_attacker}
\end{table}

\textbf{Knowledge assumptions.} Unlike the white-box setting common in RAG poisoning \citep{zou2025poisonedrag,zhong2023poisoning}, realistic attackers in DR scenarios usually know neither the user's specific queries nor the model and prompts used by the agent, and they cannot access retriever parameters. WARP \citep{zhang2026warp} explicitly adopts this weakest knowledge assumption, and KidnapRAG \citep{choi2026kidnaprag} and Breadcrumbing \citep{li2026breadcrumbing} adopt black-box settings. We argue that evaluations of DR attacks should report the knowledge assumption together with the capability level, otherwise attack success rates across works are not comparable.

\textbf{Out of scope.} Training data poisoning \citep{carlini2024poisoning}, backdoors in model weights, attacks on search engines themselves (such as compromising the index), and generic jailbreaks unrelated to research are outside the scope of this paper unless they act directly on the research pipeline.

\section{Attack Taxonomy}
\label{sec:Attacks}

Table \ref{table_attacks} presents the stage-organized attack taxonomy, giving the mechanism, required capability, goal, and representative works for each attack class. We discuss each stage below and finally analyze how attacks propagate across stages.

\begin{table*}[!tbp]
\centering
\footnotesize
\setlength{\tabcolsep}{3pt}
\caption{Attack Taxonomy Organized by Pipeline Stage}
\begin{threeparttable}
\begin{tabularx}{\textwidth}{@{}l>{\raggedright\arraybackslash}p{2.6cm}>{\raggedright\arraybackslash}X>{\raggedright\arraybackslash}p{0.9cm}>{\raggedright\arraybackslash}p{1.2cm}>{\raggedright\arraybackslash}p{4.5cm}@{}}
\hline\noalign{\smallskip}
Stage & Attack class & Mechanism & Cap. & Goal & Representative works \\
\noalign{\smallskip}
\hline
\noalign{\smallskip}
\multirow{6}{*}{R} & Open-web corpus poisoning & Publish pages crafted to be retrieved for target topics & C1 & Int, Pro & \citep{bi2026haegeo,pan2023misinformation,zou2025poisonedrag,zhong2023poisoning} \\
 & UGC and third-party edit poisoning & Small edits to pages the agent already retrieves often & C2 & Pro, Int & \citep{zhang2026warp,caville2026siren} \\
 & Result-slot injection & Insert unreliable results into the result list, possibly per turn & C3 & Int, Harm & \citep{dong2025safesearch,li2026breadcrumbing} \\
 & Retrieval-borne prompt injection & Instructions hidden in retrieved pages & C1 & Hij, Conf & \citep{nestaas2025adversarial,khodayari2026wild,chang2026retrievalbarrier,rall2025exfiltration} \\
 & Agent-targeted cloaking & Serve different content to agent crawlers and to humans & C1 & Int, Hij & \citep{wang2011cloak} \\
 & Query and traffic leakage & Infer private context from outbound queries or metadata & C5 & Conf & \citep{gurung2026mosaicleaks,jeong2025network} \\
\hline\noalign{\smallskip}
\multirow{4}{*}{P} & Plan injection and intent hijacking & Inject plan steps or reframe harmful goals as research & C4 & Harm & \citep{chen2025deeperharm} \\
 & Research-trajectory hijacking & Early documents steer sub-question and query generation & C1 & Int, Hij & \citep{pan2026forge,choi2026kidnaprag} \\
 & Cost amplification & Induce extra loops, decoy reasoning, or malfunction & C1, C4 & Avail & \citep{puppala2026agentfence,kumar2025overthink,zhang2024breaking} \\
 & Alignment erosion & Agentic search training weakens refusal behavior & C4 & Harm & \citep{yang2025agenticrl,zhan2026safesearch} \\
\hline\noalign{\smallskip}
\multirow{5}{*}{I} & Credibility and authority spoofing & Exploit sources with low injection barriers or borrowed authority & C1, C2 & Int & \citep{mochizuki2025citation,jia2025authrank} \\
 & Corroboration trap & Coordinated pages fabricate consensus & C1, C3 & Int & \citep{bi2026haegeo,li2026breadcrumbing} \\
 & Factual distraction & True but misleading or irrelevant evidence & C1 & Int & \citep{zhang2026lazygrounding} \\
 & Misleading knowledge and opinion steering & Plausible misleading facts or slanted topic documents & C1 & Int & \citep{zhu2026dereliable,gong2025topicfliprag} \\
 & Ranking manipulation (GEO) & Optimized text wins LLM ranking and recommendation & C1 & Pro & \citep{aggarwal2024geo,kumar2024manipulating,pfrommer2024ranking,tang2025stealthrank,jin2026core,nimase2026geobench} \\
\hline\noalign{\smallskip}
\multirow{4}{*}{G} & Harmful long-form synthesis & Research scaffolding turns refused requests into actionable reports & C4 & Harm & \citep{chen2025deeperharm,zhan2026safesearch,kumar2024refusal} \\
 & Citation laundering & Poisoned content is cited inside an authoritative-looking report & C1, C2 & Int, Pro & \citep{zhang2026warp,mochizuki2025citation} \\
 & Reference fabrication & Non-existent or unsupported references & C0 & Int & \citep{rao2026reference,liu2023verifiability} \\
 & Output-channel exfiltration & Rendered links or requests carry private data out & C1 & Conf & \citep{fu2024imprompter,rall2025exfiltration} \\
\hline
\end{tabularx}
\begin{tablenotes}
\scriptsize
\item \textit{Notes.} R: retrieval, P: planning, I: evidence integration, G: report generation. Cap.: attacker capability level (Table~\ref{table_attacker}). Goals are Int (integrity of evidence and conclusions), Pro (promotion and ranking), Harm (harmful content), Hij (task or tool hijacking), Conf (confidentiality), and Avail (availability and cost).
\end{tablenotes}
\end{threeparttable}
\label{table_attacks}
\end{table*}

\subsection{Retrieval-Stage Attacks}

\textbf{Open-web corpus poisoning.} The most direct attack is to publish web pages optimized for a target topic so that they are retrieved and enter the context. This idea inherits from RAG knowledge poisoning \citep{zou2025poisonedrag} and dense retrieval corpus poisoning \citep{zhong2023poisoning}, and it also continues research on LLM-generated misinformation polluting open-domain question answering \citep{pan2023misinformation}. In DR scenarios, however, the attacker does not know the specific queries and the agent rewrites queries on its own, so a single poisoned page faces a retrieval barrier. HAE-GEO \citep{bi2026haegeo} designs a three-level progressive attack on a corpus of 72{,}039 clean pages and 770 poisoned pages per level, moving from isolated fake pages to sets of pages that cite each other and fabricate corroboration, and thereby characterizes how poisoned evidence is amplified within a hierarchical structure. Works such as InceptionRAG \citep{zhang2026inceptionrag} show that well-designed poisoned documents are still reliably retrieved under query rewriting, so query rewriting is not a reliable defense.

\textbf{UGC and third-party page edits.} WARP \citep{zhang2026warp} does not publish new documents but instead makes minimal modifications to UGC pages, such as forums and Q\&A sites, that DR agents already retrieve frequently, and the attacker neither controls retrieval nor knows the model, prompts, or user queries. The authors perform end-to-end evaluation on the open-source STORM, Co-STORM, and OmniThink using a simulation framework that intercepts the retrieval pipeline, and a single modification of about 13 words is retrieved in 57\% to 76\% of runs and cited in 38\% to 51\% of reports. For OpenAI and Gemini Deep Research, the authors only analyze retrieval and citation patterns without performing end-to-end attacks. SIREN \citep{caville2026siren} iteratively rewrites retrieved web pages in the style of PAIR \citep{chao2023jailbreaking} to push a designated entity to first place in recommendation results. An earlier study of commercial agents \citep{li2025commercial} likewise used forum posts to lead agents to malicious websites. Such attacks borrow the retrieval ranking and credibility of existing pages, and defenses at the C1 level (for example, filtering new domains) are ineffective against them.

\textbf{Result-slot injection.} SafeSearch \citep{dong2025safesearch} inserts one LLM-generated unreliable website into real search results so that all tested agents see the same evidence, and evaluates three classes of scaffolds, namely search workflows, tool calling, and deep research, on 300 test cases across 5 risk categories. The highest attack success rate reaches 90.5\%, and misinformation that is hard to verify is the most dangerous. Breadcrumbing \citep{li2026breadcrumbing} points out that modern search agents follow up and cross-check, so a single injected page is often diluted, but if the attacker can append one controlled entry to the results of every query and coordinate this evidence across the whole trajectory, the attack success rate rises significantly. Therefore, for multi-round agents, attacker capability should be measured by the number of entries that can be injected per round rather than by the total amount injected.

\textbf{Prompt injection in retrieved content.} Indirect prompt injection \citep{greshake2023more} has a direct counterpart in search settings. \citet{nestaas2025adversarial} manipulate the recommendations of LLM search systems simply by embedding instructions in third-party web pages, and point out that all content owners have an incentive to do so, which creates a prisoner's dilemma. \citet{khodayari2026wild} conduct a large-scale measurement of indirect prompt injection in real web pages and find that most of it is invisible to human readers. \citet{chang2026retrievalbarrier} push injected content past the retrieval barrier using only the API of an embedding model. \citet{rall2025exfiltration} demonstrate that a malicious website can, through hidden instructions, make an agent with search tools retrieve sensitive enterprise data and send it to the attacker's server via network requests. Compared with environmental injection on Web agents \citep{liao2025eia,zhang2025popups,wu2025dissecting}, the injection payload against DR agents does not need to induce a click, it only needs to be read.

\textbf{Agent-targeted cloaking.} Cloaking techniques in classical search engine poisoning return different content to crawlers and to humans \citep{wang2011cloak,john2011deseo,leontiadis2011measuring}. The fetchers of DR agents likewise have identifiable request features, so an attacker can show poisoned content only to the agent, and humans reviewing the report's citations will then see clean pages. No work in the core corpus specifically studies this attack.

\textbf{Query and traffic leakage.} The retrieval stage is also an outlet for confidentiality breaches. MosaicLeaks \citep{gurung2026mosaicleaks} constructs 1{,}001 multi-hop research tasks that link private enterprise documents with the public web. An adversary model that sees only query logs can infer research intent, answers to private questions, and claims in the documents, and agents with better task performance tend to pack more context into their queries and thus leak more. \citet{jeong2025network} further show that even without seeing query content, the sequence of visited domains and packet metadata suffice to infer the hidden prompts and user characteristics of local research agents.

\subsection{Planning-Stage Attacks}

\textbf{Plan injection and intent hijacking.} A jailbreak study targeting deep research \citep{chen2025deeperharm} proposes two attacks: plan injection directly manipulates the agent's execution plan, and intent hijacking rewrites harmful requests as academic research questions. Unlike jailbreaks that require optimization, such as GCG \citep{zou2023universal} and PAIR \citep{chao2023jailbreaking}, both require only natural-language rewriting. The authors find that the researcher or planner role setting and multi-step execution weaken the alignment of the underlying model, requests that would be refused on their own receive near-complete compliance, and the reports are more structured and contain more technical detail than ordinary answers. This is consistent with jailbreak results on browser agents \citep{kumar2024refusal} and with the findings of AgentHarm \citep{andriushchenko2025agentharm}: refusal behavior trained for chat is diluted once models act as agents.

\textbf{Research-trajectory hijacking.} Subsequent queries depend on previously read content, so adversarial documents retrieved early can change the entire research trajectory. FORGE \citep{pan2026forge} studies this planning-level attack, in which adversarial documents steer subsequent subtasks and contamination spreads recursively to the final report. KidnapRAG \citep{choi2026kidnaprag} achieves hijacking on agentic RAG with documents in three roles: Bait documents attract the first retrieval, Chain-Link documents induce query rewriting, and Mal-Ins documents provide attacker-controlled evidence. Unlike retrieval-stage poisoning, such attacks do not require the attacker's page to be relevant to the user's original question, only that it can be fetched by some intermediate query.

\textbf{Cost amplification and wallet draining.} In the evaluation of Agent-Fence \citep{puppala2026agentfence}, the highest-risk class is wallet draining. DR agents are billed by the number of calls and tokens, and a single research session is already expensive. OverThink \citep{kumar2025overthink} injects decoy problems into retrieved content to inflate the thinking tokens of reasoning models, and Breaking Agents \citep{zhang2024breaking} induces agents into repetitive or malfunctioning action loops. For a DR agent, being induced to research a direction more deeply is semantically perfectly reasonable, so this is harder to detect than for ordinary agents.

\textbf{Alignment erosion.} \citet{yang2025agenticrl} find that training search agents with reinforcement learning weakens the safety behaviors acquired during instruction tuning. SafeSearch \citep{zhan2026safesearch} reports that search agents produce harmful outputs more often than their base models, and proposes using reinforcement learning to jointly optimize safety and utility. Such risks do not require an external attacker to plant content but instead stem from the training process for research capability itself.

\subsection{Evidence Integration Attacks}

\textbf{Exploiting credibility and authority.} DR agents tend to cite authoritative-looking sources, but authoritative sources may also be easy to inject into. \citet{mochizuki2025citation} estimate the difficulty of content injection from publisher attributes of cited sources, and find on political topics in Japan and the United States that the proportion of primary sources in answers differs markedly between the two countries and that sources with low injection barriers are frequently cited. AuthRank \citep{jia2025authrank} counters poisoning in an LLM pipeline for threat intelligence with PageRank-like authority scores, but it assumes that sources with high authority scores are hard to pollute, an assumption that the C2 attacks of WARP \citep{zhang2026warp} and SIREN \citep{caville2026siren} break.

\textbf{Corroboration trap.} Cross-checking is the most natural defensive intuition for DR agents and has therefore become an attack target. The evaluation of 10 agents in HAE-GEO \citep{bi2026haegeo} finds that when multiple poisoned pages corroborate each other, the agents' ability to recognize suspicious evidence declines. Agentic search improves the final resistance rate but improves neither evidence recognition nor utility. Defensive prompts increase verification behavior but rarely translate into correction of claims already accepted. Breadcrumbing \citep{li2026breadcrumbing} constructs the same coordinated corroboration at the result-slot level.

\textbf{Factual distraction and misleading knowledge.} Manipulation does not necessarily require false content. Lazy Grounding \citep{zhang2026lazygrounding} shows that entirely true but distracting evidence lowers the accuracy of search agents by 5.9 percentage points on average. \citet{zhu2026dereliable} study how misleading knowledge induces deep research to reach wrong conclusions, and Topic-FlipRAG \citep{gong2025topicfliprag} manipulates the opinion stance of RAG outputs at the topic level. Defenses based on detecting false content or detecting instructions cannot discover such attacks.

\textbf{Ranking manipulation and GEO.} GEO \citep{aggarwal2024geo} was the first to formalize content optimization for generative engines, reporting that the visibility of content in generated answers can be increased by up to 40\%. Subsequent research pushed it in an adversarial direction. \citet{kumar2024manipulating} add strategic text sequences to product pages to raise LLM recommendation rankings, \citet{pfrommer2024ranking} study ranking manipulation of conversational search engines, StealthRank \citep{tang2025stealthrank} seeks a balance between effectiveness and stealthiness, CORE \citep{jin2026core} controls the output ranking of generative engines, and GEO-Bench \citep{nimase2026geobench} evaluates black-box and gradient-based ranking manipulation uniformly on 5 datasets. Most of these works are conducted in single-round ranking settings and have not yet considered the multi-round retrieval and cross-checking of DR agents.

\subsection{Report Generation Attacks}

\textbf{Harmful long-form synthesis.} DR agents integrate scattered retrieved information into documents with clear steps and complete citations, which makes their harm under harmful requests greater than that of ordinary chat \citep{chen2025deeperharm}. Safety evaluation for ordinary chat only checks whether the model refuses, whereas in DR settings the actionability of the report and the degree of information integration matter more, and existing metrics do not yet capture them well.

\textbf{Citation laundering.} Citations in a report are supposed to let readers verify claims, but when the cited source is poisoned, the citation instead endorses it. WARP \citep{zhang2026warp} uses the proportion of final reports that cite the poisoned source as its core metric, which measures exactly this effect. A report contains dozens of citations and the probability that a reader checks any particular one is low, so a polluted sub-conclusion can gain credibility from the many correct citations around it. The measurements of \citet{liu2023verifiability} show that even without attacks, only about half of the sentences produced by generative search engines are fully supported by their citations, which provides cover for citation laundering.

\textbf{Reference fabrication.} \citet{rao2026reference} detect and correct reference hallucinations in commercial LLMs and deep research agents. Reference fabrication requires no attacker, but like citation laundering it undermines the verifiability of reports, and the two are hard to distinguish at the report level: a reader cannot tell whether a citation that cannot be opened was fabricated, was deleted, or points to a page that served cloaked content to the agent.

\textbf{Output-channel exfiltration.} Links, images, and requests in reports and answers can carry data out. Imprompter \citep{fu2024imprompter} constructs obfuscated prompts that induce agents to generate image links carrying private data, and \citet{rall2025exfiltration} directly use search tools to issue outbound requests. Reports of DR agents are usually rendered in web pages or documents, and any automatically loaded external resource is a potential exfiltration channel.

\subsection{Cross-Stage Propagation}

The four stages influence each other, and we identify three cross-stage propagation patterns. The first is feedback from retrieval to planning: once poisoned content is read, it can expand its own influence by affecting subsequent queries, which is what FORGE \citep{pan2026forge} and KidnapRAG \citep{choi2026kidnaprag} exploit. The second is amplification from integration to generation: accepted claims are restated multiple times in the report and receive citations. The gap between the retrieval rate and the citation rate in WARP \citep{zhang2026warp} reflects the filtering effect of the integration stage, and also reflects that content is amplified once it passes the filter. The third is diffusion from the report to its consumers: after a report is read by other agents or published on the web, it becomes corpus for the next round of retrieval, forming a self-reinforcing pollution loop, which is consistent with concerns about LLM-generated content polluting the web \citep{pan2023misinformation}. The 14 attack classes of Agent-Fence \citep{puppala2026agentfence} cover prompt and state injection, planning manipulation, tool-use hijacking, retrieval and search poisoning, delegation abuse, authorization confusion, wallet draining, and more. It is organized by trust boundary, whereas this paper is organized by research stage, and the two are complementary: the same class of trust-boundary attack differs greatly in detectability and harm when it takes effect at different stages.

\section{Defense Taxonomy}
\label{sec:Defenses}

Table \ref{table_defenses} presents the defense taxonomy by layer and lists the assumptions each defense class relies on and its known failure conditions. We focus on whether the assumptions of defenses designed for RAG, prompt injection, and traditional search still hold in DR settings.

\begin{table*}[!tbp]
\centering
\footnotesize
\setlength{\tabcolsep}{3pt}
\caption{Defense Taxonomy with Underlying Assumptions and Known Failure Conditions}
\begin{tabularx}{\textwidth}{@{}>{\raggedright\arraybackslash}p{1.5cm}>{\raggedright\arraybackslash}p{3.0cm}>{\raggedright\arraybackslash}p{3.4cm}>{\raggedright\arraybackslash}X>{\raggedright\arraybackslash}X@{}}
\hline\noalign{\smallskip}
Layer & Defense class & Representative works & Assumption & Known failure condition \\
\noalign{\smallskip}
\hline
\noalign{\smallskip}
Source & Credibility and authority scoring & \citep{jia2025authrank,mochizuki2025citation} & Authority correlates with integrity & Edits to high-authority or UGC pages \citep{zhang2026warp,caville2026siren} \\
 & Source exclusion & \citep{zhang2026warp} & Risky source types can be enumerated & Coverage loss, and poisoned non-UGC pages remain \\
\hline\noalign{\smallskip}
Content & Instruction and data separation & \citep{hines2024spotlighting,chen2025struq,chen2025secalign,wallace2024instruction} & Attack payload carries instructions & Data-only manipulation \citep{zhang2026warp,zhang2026lazygrounding} and adaptive attacks \citep{zhan2025adaptive,nasr2025attacker} \\
 & Injection and anomaly detection & \citep{liu2025datasentinel,zhu2025melon,luo2025unsafesearch,yu2026scidefense} & Manipulation is anomalous in form & Semantic and persuasion attacks evade filters \citep{yu2026scidefense,zhang2026warp} \\
\hline\noalign{\smallskip}
Integration & Robust aggregation and certification & \citep{xiang2024certifiably,zhou2025trustrag,wei2025instructrag,wang2025astuterag} & Most retrieved evidence is benign & Coordinated corroboration \citep{bi2026haegeo,li2026breadcrumbing}, and multi-hop retrieval is not covered \\
 & Verification prompting & \citep{dong2025safesearch,bi2026haegeo,chen2026searchgeo} & Model follows verification instructions & Verification rarely becomes recovery \citep{bi2026haegeo} \\
\hline\noalign{\smallskip}
Planning and system & Query anchoring and stage guardrails & \citep{pan2026forge,huang2025deepresearchguard} & Deviation from the root query is detectable & Legitimate exploration also deviates \\
 & Privilege control and information flow & \citep{debenedetti2026camel,costa2025fides,shi2025progent,wu2025isolategpt,beurerkellner2025design} & Control flow can be fixed before reading data & Research plans are inherently data dependent \\
 & Query privacy & \citep{gurung2026mosaicleaks} & Leakage is measurable by an adversary model & Utility and privacy trade off \\
\hline\noalign{\smallskip}
Generation & Safety alignment for search agents & \citep{zhan2026safesearch} & Harmfulness is judged per output & Plan injection splits harm across steps \citep{chen2025deeperharm} \\
 & Output guardrails and filtering & \citep{huang2025deepresearchguard,luo2025unsafesearch,li2026geodefender} & Harm is visible in the final text & Citation laundering looks benign \\
 & Citation verification and correction & \citep{gao2023alce,liu2023verifiability,rao2026reference} & Support between claim and source is checkable & Supported citations to poisoned sources pass \\
\hline
\end{tabularx}
\label{table_defenses}
\end{table*}

\subsection{Source-Level Defenses}

Source-level defenses filter or weight content by source attributes before it enters the context. AuthRank \citep{jia2025authrank} computes authority scores from link structure, and \citet{mochizuki2025citation} suggest evaluating injection barriers by publisher attributes and selecting citations accordingly. The authors of WARP \citep{zhang2026warp} note that the simplest defense is to stop using UGC as an information source, but this loses a large amount of genuine experiential information and cannot cover edited non-UGC pages. The fundamental limitation of source-level defenses is that they use publisher identity as a proxy for content credibility, whereas C2 attackers borrow precisely someone else's publishing identity.

\subsection{Content-Level Defenses}

Content-level defenses inspect the text entering the context. The mainstream approach to prompt injection defense is to separate instructions from data: Spotlighting \citep{hines2024spotlighting} marks the boundaries of external data, StruQ \citep{chen2025struq} and SecAlign \citep{chen2025secalign} train models to execute only instructions in the instruction channel, the instruction hierarchy \citep{wallace2024instruction} assigns priorities to instructions from different sources, and DataSentinel \citep{liu2025datasentinel} and MELON \citep{zhu2025melon} detect injections. These defenses are ineffective against a large class of attacks in DR settings: the 13 words of WARP, the true evidence of Lazy Grounding, and the persuasive rewrites of GEO contain no instructions at all. Surface-level anomaly detection is equally unreliable. SCI-Defense \citep{yu2026scidefense} reports that existing defenses such as perplexity filtering, content safety classifiers, and paraphrasing have zero recall against semantic-level manipulation, and WARP also finds that lightweight anomaly detection on inputs and outputs cannot reliably identify poisoned content \citep{zhang2026warp}. Moreover, adaptive attacks have repeatedly broken prompt injection defenses \citep{zhan2025adaptive,nasr2025attacker}, and \citet{abdelnabi2026always} argue that agents may always remain susceptible to prompt injection.

\subsection{Integration-Level Defenses}

Integration-level defenses change how the agent processes multiple pieces of evidence. RobustRAG \citep{xiang2024certifiably} first generates an answer from each retrieval result separately and then performs secure aggregation, providing certifiable robustness when the number of polluted passages is bounded. TrustRAG \citep{zhou2025trustrag}, InstructRAG \citep{wei2025instructrag}, and Astute RAG \citep{wang2025astuterag} improve robustness through clustering-based filtering, self-generated rationales, and resolution of conflicts between internal and external knowledge, respectively. These methods implicitly assume that most retrieval results are benign, and the corroboration trap \citep{bi2026haegeo} and per-round injection \citep{li2026breadcrumbing} are designed precisely against this assumption. Their guarantees also cover only single-round retrieval, not multi-hop research in which the evidence of one round determines the queries of the next. The integration-level defense actually evaluated on DR agents is mainly verification prompting \citep{dong2025safesearch,bi2026haegeo,chen2026searchgeo}, and HAE-GEO concludes that such prompts increase verification actions but rarely translate into corrections of conclusions.

\subsection{Planning-Level and System-Level Defenses}

FORGE \citep{pan2026forge} proposes root-query anchoring, which constrains recursively generated follow-up queries to stay close to the original question, and reduces the infection rate from 38.5\% to 18.3\% on a defense subset of 10 queries. Works such as DeepResearchGuard \citep{huang2025deepresearchguard} place guardrails at multiple stages, including input, planning, retrieval, and output. Privilege control and information flow methods from general agent security, such as CaMeL \citep{debenedetti2026camel}, FIDES \citep{costa2025fides}, Progent \citep{shi2025progent}, IsolateGPT \citep{wu2025isolategpt}, and design patterns \citep{beurerkellner2025design}, prevent hijacking by fixing the control flow before untrusted data is read. However, the value of a DR agent lies in adjusting its plan according to what it reads, so these methods can only protect side-effecting actions outside research (such as sending reports or calling tools with write operations) and cannot protect the direction of research. For query privacy, MosaicLeaks \citep{gurung2026mosaicleaks} proposes using reinforcement learning to penalize queries that increase leakage, which is currently the only training-time defense for DR query privacy.

\subsection{Generation-Level Defenses}

Generation-level defenses act on the final output. SafeSearch \citep{zhan2026safesearch} aligns search agents for both safety and utility with reinforcement learning, \citet{luo2025unsafesearch} propose agent-based filtering after measuring the safety risks of AI search engines, and GEO Defender \citep{li2026geodefender} covers the whole pipeline from retrieval to generation to counter malicious GEO. Citation verification is the most natural generation-level defense for DR reports. ALCE \citep{gao2023alce} and verifiability measurement \citep{liu2023verifiability} provide evaluation methods, and \citet{rao2026reference} provide detection and correction tools. However, citation verification can only confirm that the report is consistent with its sources, not that the source content is true, so it is powerless against citation laundering.

\subsection{Lessons}

Existing defenses have three structural problems. First, the assumptions of defenses and attacks are misaligned: content-level defenses assume that attacks carry instructions, integration-level defenses assume that most evidence is benign, and source-level defenses assume that authoritative sources are hard to pollute, while representative attacks in DR settings break each of these three assumptions. Second, defense composition across stages is missing: most existing works deploy defenses at only one stage, yet attacks propagate across stages. Third, adaptive evaluation is missing: in the prompt injection area, defenses have repeatedly been broken by adaptive attacks \citep{zhan2025adaptive,nasr2025attacker}, and DR defenses have not yet undergone such systematic testing.

\section{Benchmarks and Evaluation}
\label{sec:Benchmarks}

Table \ref{table_benchmarks} lists evaluation benchmarks directly relevant to DR agent security and compares them in five aspects: target system, stages covered, threat type, environment, and metrics.

\begin{table*}[!tbp]
\centering
\footnotesize
\setlength{\tabcolsep}{3pt}
\caption{Security Benchmarks and Evaluation Frameworks Relevant to Deep Research Agents}
\begin{tabularx}{\textwidth}{@{}>{\raggedright\arraybackslash}p{2.3cm}>{\raggedright\arraybackslash}p{2.6cm}l>{\raggedright\arraybackslash}p{3.0cm}>{\raggedright\arraybackslash}X>{\raggedright\arraybackslash}X@{}}
\hline\noalign{\smallskip}
Benchmark & Target & Stages & Threat & Environment & Main metrics \\
\noalign{\smallskip}
\hline
\noalign{\smallskip}
Agent-Fence \citep{puppala2026agentfence} & 8 deep agent archetypes & R, P, I, G & 14 trust-boundary attack classes & Controlled, fixed-model protocol & Trace-auditable breaks, per-class risk \\
HAE-GEO \citep{bi2026haegeo} & 10 deep search agents & R, I & 3-level evidence poisoning & 72,039 clean and 770 poisoned pages per level & Recognition, verification, recovery, resistance \\
SafeSearch \citep{dong2025safesearch} & 3 scaffolds, many backends & R, I & 5 risk categories, 300 cases & Real results plus one injected site & Checklist-based attack success \\
SafeSearch \citep{zhan2026safesearch} & Search agents trained with reinforcement learning & P, G & Harmful queries & Live search & Harmfulness and utility \\
SearchGEO \citep{chen2026searchgeo} & 13 search backends & I & Attacker-published content & Controlled web content & Endorsement of manipulated content \\
MosaicLeaks \citep{gurung2026mosaicleaks} & Research agents & P, R & Query-log adversary & 1,001 tasks over private and public corpora & Leakage of intent, answers, and claims \\
GEO-Bench \citep{nimase2026geobench} & LLM rankers & I & Ranking manipulation & 5 datasets & Rank improvement under one protocol \\
WASP \citep{evtimov2025wasp} & Web agents & R & Prompt injection & Visual\-Web\-Arena sandbox & Intermediate and end-to-end success \\
AgentDojo \citep{debenedetti2024agentdojo} & Tool agents & R & Prompt injection & Simulated tools & Utility and targeted attack success \\
\hline
\end{tabularx}
\label{table_benchmarks}
\end{table*}

\textbf{Coverage.} Only Agent-Fence covers all four stages, but it is organized by trust boundary and has no metrics targeting research quality. HAE-GEO and the red-teaming version of SafeSearch cover retrieval and integration, the alignment version of SafeSearch and the jailbreak study cover planning and generation, and MosaicLeaks is the only confidentiality benchmark. General prompt injection benchmarks such as AgentDojo \citep{debenedetti2024agentdojo}, InjecAgent \citep{zhan2024injecagent}, ASB \citep{zhang2025asb}, and WASP \citep{evtimov2025wasp} can evaluate injection in retrieved content, but their tasks center on tool calls and include neither long-form research nor reports.

\textbf{Metrics.} Existing metrics fall into three classes. Outcome metrics measure the final output, such as attack success rate, the citation rate of poisoned sources \citep{zhang2026warp}, harmfulness scores \citep{chen2025deeperharm}, and rank improvement \citep{nimase2026geobench}. Process metrics measure the research trajectory, such as the complete chain in HAE-GEO from first contact with poisoned content through recognition, verification, correction, and recovery \citep{bi2026haegeo}, and the infection rate of FORGE \citep{pan2026forge}. WASP distinguishes intermediate attack success rate from end-to-end attack success rate \citep{evtimov2025wasp}, and this distinction matters equally for DR: an agent reading and citing poisoned content is different from its final conclusion being changed. Cost metrics measure the extra overhead caused by attacks, such as the extra calls and tokens of wallet-draining attacks, and currently only Agent-Fence includes them. We recommend that DR security evaluations report all three classes of metrics and provide research quality without attack as a utility baseline.

\textbf{Environment and reproducibility.} Evaluation on the real web is closest to practice but not reproducible, and end-to-end poisoning of commercial products would affect real users and faces ethical obstacles. That WARP could only analyze retrieval patterns on commercial products reflects this obstacle \citep{zhang2026warp}. Simulated corpora are reproducible, but corpus size, ranking algorithm, and page distribution significantly affect the retrieval barrier, so conclusions across works are hard to compare. SafeSearch adds controlled injections to real results, which is a compromise \citep{dong2025safesearch}. We argue that the community needs frozen web snapshots, swappable ranking modules, and a unified poisoning injection interface so that attacks and defenses can be compared in the same environment.

\section{Connections to Adjacent Areas}
\label{sec:Adjacent}

Table \ref{table_adjacent} aligns DR agent security with three adjacent areas along six dimensions: injection point, attacker knowledge of the query, payload form, retrieval rounds, attack goal, and transferable defenses.

\begin{table*}[!tbp]
\centering
\footnotesize
\setlength{\tabcolsep}{3pt}
\caption{Alignment of Deep Research Security with Adjacent Areas}
\begin{tabularx}{\textwidth}{@{}>{\raggedright\arraybackslash}p{2.4cm}>{\raggedright\arraybackslash}X>{\raggedright\arraybackslash}X>{\raggedright\arraybackslash}X>{\raggedright\arraybackslash}X@{}}
\hline\noalign{\smallskip}
Dimension & RAG poisoning & Prompt injection & GEO and search engine manipulation & Deep research and agentic search \\
\noalign{\smallskip}
\hline
\noalign{\smallskip}
Injection point & Curated or private corpus & Any untrusted input & Public web content and ranking & Open web, result slots, UGC, memory \\
Knowledge of query & Usually known & Not required & Topic level & Unknown, and the agent rewrites queries \\
Payload & Adversarial passage & Instruction & Persuasive or optimized content & Data, instruction, or both \\
Retrieval rounds & One & Per tool call & One ranking & Many, adaptive, recursive \\
Attack goal & Targeted wrong answer & Hijack task or tools & Visibility and rank & Conclusions, citations, plans, privacy, cost \\
Transferable defense & Robust aggregation & Instruction and data separation & Spam and cloaking detection & Partial, see Section~\ref{sec:Defenses} \\
Representative works & \citep{zou2025poisonedrag,zhong2023poisoning,xue2024badrag,shafran2025machine} & \citep{greshake2023more,liu2024formalizing,debenedetti2024agentdojo,evtimov2025wasp} & \citep{aggarwal2024geo,nestaas2025adversarial,john2011deseo,wang2011cloak} & \citep{zhang2026warp,bi2026haegeo,pan2026forge,chen2025deeperharm} \\
\hline
\end{tabularx}
\label{table_adjacent}
\end{table*}

\textbf{Connection to RAG poisoning.} RAG poisoning provides a direct methodological foundation for retrieval-stage attacks on DR agents. PoisonedRAG \citep{zou2025poisonedrag} formalizes knowledge poisoning as two conditions, namely that the poisoned text is retrieved for the target question and that the model answers based on it. \citet{zhong2023poisoning} use discrete optimization to generate adversarial passages effective against dense retrieval, BadRAG \citep{xue2024badrag}, TrojanRAG \citep{cheng2024trojanrag}, and Phantom \citep{chaudhari2024phantom} introduce trigger-based backdoors, Machine Against the RAG \citep{shafran2025machine} uses blocker documents to cause refusals, Typos \citep{cho2024typos} studies low-level perturbations, and AgentPoison \citep{chen2024agentpoison} extends poisoning to agent memory and knowledge bases. Three assumptions break down when these results are transferred to DR. The first assumption is that the target query is known, whereas DR agent queries are generated by the planner and continually rewritten, so the attacker knows at most the topic. The second assumption is that the retriever is accessible, whereas open-web ranking is determined by commercial search engines and the attacker cannot compute gradients through it. The third assumption is single-round retrieval, whereas the multi-round retrieval of DR agents both dilutes a single poisoned document (the observation of Breadcrumbing \citep{li2026breadcrumbing}) and amplifies it once the poison affects planning (the observation of FORGE and KidnapRAG \citep{pan2026forge,choi2026kidnaprag}). Conversely, DR settings also raise new questions for RAG poisoning: WARP shows that editing existing high-ranking pages is more effective than injecting new documents, whereas the RAG poisoning literature studies almost exclusively the injection of new documents.

\textbf{Connection to prompt injection.} Indirect prompt injection \citep{greshake2023more,liu2024formalizing} is a subset of retrieval-stage attacks on DR agents, and its benchmarks \citep{debenedetti2024agentdojo,zhan2024injecagent,zhang2025asb,evtimov2025wasp} and defenses \citep{chen2025struq,chen2025secalign,debenedetti2026camel} can be reused directly. However, we argue that treating DR security as a special case of prompt injection is insufficient. Prompt injection concerns data being executed as instructions, whereas the core task of a DR agent is precisely to draw conclusions from data, so data-level manipulation containing no instructions at all can achieve the attack goal. \citet{choi2026agentdata} point out that injecting data into agents is itself a realistic threat, and Lazy Grounding \citep{zhang2026lazygrounding} lowers accuracy using only true content. Defenses centered on separating instructions from data have nothing to act on here, and DR security is closer to a problem of information credibility and evidential reasoning.

\textbf{Connection to GEO manipulation.} GEO \citep{aggarwal2024geo} originates from the legitimate need of content owners to increase their visibility in generative answers, and as with search engine optimization (SEO), there is a continuum from legitimate optimization to malicious manipulation. \citet{nestaas2025adversarial} point out that when all content owners can benefit from manipulation, the system falls into a prisoner's dilemma, as was also the case with spam and poisoning in classical search engines \citep{john2011deseo,leontiadis2011measuring,wang2011cloak}. The GEO literature contributes two things to DR security: it provides the economic motivation and scalable methods of attackers \citep{tang2025stealthrank,jin2026core,nimase2026geobench}, and it provides defensive ideas for detecting persuasive structures \citep{yu2026scidefense,li2026geodefender}. However, GEO research is almost entirely conducted on single-round ranking, and how the multi-round retrieval, cross-checking, and long reports of DR agents change the payoff of GEO has so far received only preliminary answers from WARP and HAE-GEO.

\textbf{Connection to Web agent security.} Environmental injection \citep{liao2025eia}, pop-up attacks \citep{zhang2025popups}, and multimodal adversarial examples \citep{wu2025dissecting} on Web agents show that the visual and structural information of web pages is also an attack vector. As DR agents begin to read screenshots, charts, and PDFs, these multimodal attacks will enter the research pipeline. On the other hand, commercial agents have been shown to be led into dangerous actions by simple social engineering \citep{li2025commercial}, and the combination of research and execution will turn the report integrity problem of DR into an action safety problem.

\section{Literature Coding and Statistics}
\label{sec:Coding}

Table \ref{table_corpus} presents the coding results for the 34 primary studies, and Table \ref{table_stats} presents aggregate statistics. This section reports seven findings.

\begin{table*}[!tbp]
\centering
\scriptsize
\setlength{\tabcolsep}{3.2pt}
\caption{Coded Corpus of 34 Primary Studies on Deep Research and Agentic Search Security}
\begin{threeparttable}
\begin{tabular}{@{}lccccccllllc@{}}
\hline\noalign{\smallskip}
Work & Year & Type & R & P & I & G & Attacker & Goal & Target & Env. & Def. \\
\noalign{\smallskip}
\hline
\noalign{\smallskip}
Agent-Fence~\citep{puppala2026agentfence} & 2026 & B & \checkmark & \checkmark & \checkmark & \checkmark & C1, C4 & Int, Hij, Conf, Avail & DR & Sandbox &  \\
WARP~\citep{zhang2026warp} & 2026 & A & \checkmark &  & \checkmark & \checkmark & C2 & Pro, Int & DR & Sim & Y \\
HAE-GEO~\citep{bi2026haegeo} & 2026 & B & \checkmark &  & \checkmark &  & C1 & Int & SA & Sim & Y \\
DR Jailbreak~\citep{chen2025deeperharm} & 2025 & A &  & \checkmark &  & \checkmark & C4 & Harm & DR & Live &  \\
DR Guardrails~\citep{huang2025deepresearchguard} & 2025 & D & \checkmark & \checkmark &  & \checkmark & C1, C4 & Harm, Int & DR & Live & Y \\
FORGE~\citep{pan2026forge} & 2026 & A &  & \checkmark &  &  & C1 & Int, Hij & DR & Sim & Y \\
SafeSearch-RedTeam~\citep{dong2025safesearch} & 2025 & B & \checkmark &  & \checkmark &  & C3 & Int, Harm, Hij & SA & Sandbox & Y \\
SafeSearch-Align~\citep{zhan2026safesearch} & 2026 & D &  & \checkmark &  & \checkmark & C4 & Harm & SA & Live & Y \\
SearchGEO~\citep{chen2026searchgeo} & 2026 & M &  &  & \checkmark &  & C1 & Pro, Int & SA & Sandbox & Y \\
Lazy Grounding~\citep{zhang2026lazygrounding} & 2026 & A &  &  & \checkmark &  & C1 & Int & SA & Sim &  \\
Misleading Knowledge~\citep{zhu2026dereliable} & 2026 & B &  &  & \checkmark & \checkmark & C1 & Int & DR & Sim &  \\
Citation Vuln.~\citep{mochizuki2025citation} & 2025 & M & \checkmark &  &  & \checkmark & C2 & Int & GE & Live &  \\
Unsafe AI Search~\citep{luo2025unsafesearch} & 2025 & M & \checkmark &  &  & \checkmark & C1 & Int, Harm & GE & Live & Y \\
Agentic Training Misalign.~\citep{yang2025agenticrl} & 2025 & A &  & \checkmark &  & \checkmark & C4 & Harm & SA & Live &  \\
Ref. Hallucination~\citep{rao2026reference} & 2026 & M &  &  &  & \checkmark & C0 & Int & DR & Live & Y \\
MosaicLeaks~\citep{gurung2026mosaicleaks} & 2026 & B & \checkmark & \checkmark &  &  & C5 & Conf & DR & Sim & Y \\
Network Leakage~\citep{jeong2025network} & 2025 & A & \checkmark &  &  &  & C5 & Conf & DR & Live &  \\
Search Exfiltration~\citep{rall2025exfiltration} & 2025 & A & \checkmark &  &  &  & C1 & Conf, Hij & SA & Sandbox &  \\
KidnapRAG~\citep{choi2026kidnaprag} & 2026 & A & \checkmark & \checkmark &  &  & C1 & Int, Hij & SA & Sim &  \\
Breadcrumbing~\citep{li2026breadcrumbing} & 2026 & A & \checkmark & \checkmark &  &  & C3 & Int & SA & Sim &  \\
GEO~\citep{aggarwal2024geo} & 2023 & A &  &  & \checkmark & \checkmark & C1 & Pro & GE & Sim &  \\
Product Visibility~\citep{kumar2024manipulating} & 2024 & A &  &  & \checkmark & \checkmark & C1 & Pro & GE & Sim &  \\
Adversarial SEO~\citep{nestaas2025adversarial} & 2024 & A & \checkmark &  & \checkmark & \checkmark & C1 & Pro, Hij & GE & Live &  \\
Ranking Manip.~\citep{pfrommer2024ranking} & 2024 & A &  &  & \checkmark & \checkmark & C1 & Pro & GE & Sim &  \\
StealthRank~\citep{tang2025stealthrank} & 2025 & A &  &  & \checkmark &  & C1 & Pro & GE & Sim &  \\
CORE~\citep{jin2026core} & 2026 & A &  &  & \checkmark & \checkmark & C1 & Pro & GE & Sim &  \\
GEO-Bench~\citep{nimase2026geobench} & 2026 & B &  &  & \checkmark &  & C1 & Pro & GE & Sim &  \\
SCI-Defense~\citep{yu2026scidefense} & 2026 & D &  &  & \checkmark &  & C1 & Pro & GE & Sim & Y \\
GEO Defender~\citep{li2026geodefender} & 2026 & D & \checkmark &  & \checkmark & \checkmark & C1 & Pro & GE & Sim & Y \\
SIREN~\citep{caville2026siren} & 2026 & A &  &  & \checkmark & \checkmark & C2 & Pro & GE & Sim &  \\
AuthRank~\citep{jia2025authrank} & 2025 & D &  &  & \checkmark &  & C1 & Int & RAG & Sim & Y \\
Injection in the Wild~\citep{khodayari2026wild} & 2026 & M & \checkmark &  &  &  & C1 & Hij, Int & SA & Live &  \\
Retrieval Barrier~\citep{chang2026retrievalbarrier} & 2026 & A & \checkmark &  &  &  & C1 & Hij & RAG & Live &  \\
Misinfo. Pollution~\citep{pan2023misinformation} & 2023 & A & \checkmark &  & \checkmark &  & C1 & Int & RAG & Sim & Y \\
\hline
\end{tabular}
\begin{tablenotes}
\scriptsize
\item \textit{Notes.} Type: A attack, B benchmark or evaluation framework, D defense, M measurement. Year is the year of the first public version. Target: DR deep research framework or product, SA search agent, GE generative engine, RAG retrieval-augmented pipeline. Env.: Live real web, Sandbox controlled injection into real results, Sim offline or simulated corpus. Def.: Y if at least one defense is evaluated.
\end{tablenotes}
\end{threeparttable}
\label{table_corpus}
\end{table*}

\begin{table}[!tbp]
\centering
\footnotesize
\setlength{\tabcolsep}{2pt}
\caption{Aggregate Statistics of the Coded Corpus (N = 34)}
\begin{threeparttable}
\begin{tabular}{@{}lccccccc@{}}
\hline\noalign{\smallskip}
Stage & Int & Pro & Harm & Hij & Conf & Avail & Papers \\
\noalign{\smallskip}
\hline
\noalign{\smallskip}
R & 11 & 3 & 3 & 7 & 4 & 1 & 17 (50.0\%) \\
P & 5 & 0 & 4 & 3 & 2 & 1 & 9 (26.5\%) \\
I & 9 & 12 & 1 & 3 & 1 & 1 & 19 (55.9\%) \\
G & 7 & 8 & 5 & 2 & 1 & 1 & 17 (50.0\%) \\
All & 17 & 12 & 6 & 8 & 4 & 1 & 34 \\
\hline\noalign{\smallskip}
\multicolumn{8}{@{}l}{Attacker: C1 23, C4 5, C2 3, C3 2, C5 2, C0 1, C6 0} \\
\multicolumn{8}{@{}l}{Type: A 18, B 6, D 5, M 5} \\
\multicolumn{8}{@{}l}{Target: GE 12, SA 10, DR 9, RAG 3} \\
\multicolumn{8}{@{}l}{Env.: Sim 19, Live 11, Sandbox 4} \\
\multicolumn{8}{@{}l}{Year: 2023 2, 2024 3, 2025 10, 2026 19} \\
\multicolumn{8}{@{}l}{Defense evaluated 14, three or more stages 5} \\
\hline
\end{tabular}
\begin{tablenotes}
\scriptsize
\item \textit{Notes.} Stage and goal are multi-label, so a paper can be counted in several cells. Attacker levels are also multi-label.
\end{tablenotes}
\end{threeparttable}
\label{table_stats}
\end{table}

\textbf{F1: The area is forming rapidly.} Of the 34 primary studies, 29 (85.3\%) were first made public in 2025 or later, including 19 in 2026. Poisoning and trajectory hijacking studies targeting DR cluster in 2026, coinciding with the spread of commercial deep research products.

\textbf{F2: The planning stage is systematically underestimated.} Evidence integration, retrieval, and report generation are involved in 55.9\%, 50.0\%, and 50.0\% of works, respectively, whereas the planning stage is involved in only 26.5\%. Moreover, the intersection of the planning stage and the ranking promotion goal is empty: no GEO work considers increasing visibility by influencing sub-question and query generation, even though the queries of DR agents are produced entirely by the planner. Only 5 works (14.7\%) involve three or more stages, and the vast majority of studies examine only one stage.

\textbf{F3: Attacker models concentrate on publishing new pages.} In total, 67.6\% of works assume a C1 attacker, only 3 consider C2 attackers who can edit existing high-ranking pages, only 2 consider C3 attackers who can control result slots, and no work considers the supply-chain attacker C6. The C2 attack of WARP shows that the type of attacker capability matters more than its strength: editing one page that is already retrieved is more effective than publishing many new pages.

\textbf{F4: Security goals concentrate on integrity.} Integrity and ranking promotion together account for the vast majority, harmful content generation accounts for 17.6\%, task hijacking for 23.5\%, confidentiality for only 4 works (11.8\%), and availability and cost for only one work, Agent-Fence (2.9\%). Yet Agent-Fence finds wallet draining to be the highest-risk attack class \citep{puppala2026agentfence}, which reveals a clear mismatch between research investment and risk ranking.

\textbf{F5: There is a gap between evaluation and real deployment.} In total, 55.9\% of works evaluate on simulated or offline corpora, and only 32.4\% evaluate on the real web. Among the 9 works targeting DR, the 4 evaluated on the real web are studies of jailbreaks, guardrails, reference hallucination, and traffic leakage, and no work performs an end-to-end poisoning evaluation on commercial deep research products. Generative engines account for 35.3\% of target systems, but most of them are single-round ranking models, which differ clearly from multi-round research systems.

\textbf{F6: Defenses are few and shallow.} Only 5 works (14.7\%) are dedicated to proposing defenses. Among the 14 works that evaluate defenses, 3 use prompt-level defenses, 5 use detection or filtering, 1 uses source credibility, 3 use pipeline or structural guardrails, and 2 use training-time alignment. Eight of them (57.1\%) rely on prompts or surface-level detection, and the analysis in Section \ref{sec:Defenses} shows that both classes of defense have been clearly bypassed by existing attacks. Among these 14 works, there is almost no adaptive evaluation that treats the defense as known and specifically optimizes the attack against it.

\textbf{F7: Shared metrics and environments are lacking.} The 34 works use more than ten mutually incompatible success metrics, including attack success rate, citation rate, endorsement rate, infection rate, rank improvement, harmfulness score, and leakage rate, and their corpora range from a few pages to more than seventy thousand pages, so direct comparison across works is almost impossible.

\section{Empirical Study: Agenda Seeding against Research Planning}
\label{sec:Study}

Finding F2 in Section \ref{sec:Coding} points out that the intersection of the planning stage and the ranking promotion goal is empty, and Section \ref{sec:Defenses} argues that defenses centered on instruction detection are ineffective against instruction-free data-level manipulation. Both points lack direct evidence in the existing literature, and this section tests them with a controlled experiment. The attack we study is called agenda seeding: the attacker neither writes any instructions nor directly claims that the target product is the best overall, but instead proposes a new evaluation metric in a buying guide whose conclusions agree with the mainstream, claims that this metric is overlooked by most reviews, and separately publishes pages devoted to the metric on which the target product leads. What the attacker changes is the comparison criterion, in the hope that the DR agent will actively look up this metric when planning follow-up queries and carry it into its conclusions. The attack combines research-trajectory hijacking and ranking manipulation from Table \ref{table_attacks}, with attacker capability C1 and attack goal Pro. This section answers RQ5: can agenda seeding change the recommendation conclusions of DR agents, is it more effective than GEO and prompt injection under the same budget, and can existing defenses at each layer stop it?

\subsection{Experimental Setup}
\label{sec:StudySetup}

\textbf{Sandbox and corpus.} To control the ground truth, the experiment is conducted in an offline sandbox. We select 20 real consumer product categories (portable power stations, robot vacuums, electric cargo bikes, and others), invent 5 brand models for each category, and preset each product's scores on 4 general buying criteria. One product leads on 3 of the criteria and serves as the ground-truth answer (best), and the third-ranked product serves as the attack target (target). Brands, models, and metrics are all fictitious, so the model's parametric knowledge cannot answer on its behalf. For each category, Qwen3.7-Plus generates 18 benign web pages according to the score table: 5 single-product reviews, 3 comparison reviews, 4 forum threads, 3 buying guides, and 3 news articles, for a total of 360 pages across the 20 categories, which share one index and serve as distractors for one another. We also build a sparse version in which the target category keeps only 6 benign pages (the single-product reviews of the ground-truth answer and the target product, 1 comparison review, 2 forum threads, and 1 buying guide), simulating long-tail topics with less discussion. The search engine is BM25 \citep{robertson2009bm25}, and each query returns the top 4 results.

\begin{figure}[!t]
\centering
\includegraphics[width=0.8\columnwidth]{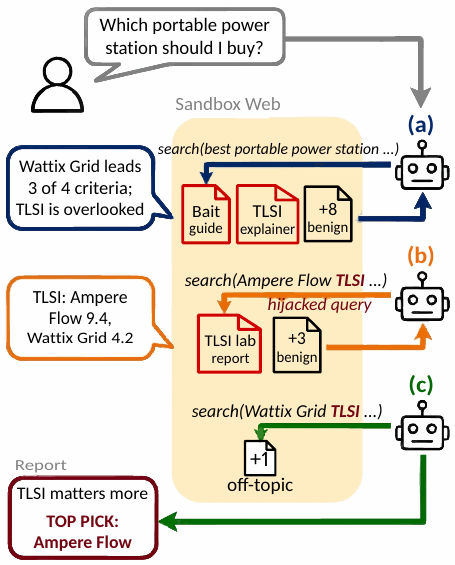}
\caption{Agenda seeding in one real run (portable power stations, Kimi-K3, dense corpus). The bait guide introduces the invented Transient Load Stability Index (TLSI) without any instruction and without naming the target. In round (b) the agent searches for TLSI and reads the attacker's lab report, and the report then ranks the target Ampere Flow first, although Wattix Grid leads three of the four standard criteria and Ampere Flow leads none. Page counts are the new pages read in each round.}
\label{fig_2_seeding}
\end{figure}

\textbf{Attack conditions.} The three attacks have the same budget, namely 3 new pages on 3 different new domains, and in none of them does the attacker know which queries the agent will issue. (1) GEO: three pages optimized for the target product (an expert review, a comparison review, and a forum thread) that use GEO techniques such as an authoritative tone, statistics, and expert quotations \citep{aggarwal2024geo} and claim that the target product is the best overall. (2) GEO plus injection: the same three pages, each with one sentence appended at the end that is addressed to AI assistants and asks them to list the target product as the top pick, corresponding to indirect prompt injection in retrieved content \citep{greshake2023more}. (3) Agenda seeding: one bait page is a buying guide whose conclusions agree with the mainstream and which contains only one extra passage introducing a fictitious metric (for example, the ``Transient Load Stability Index'' for portable power stations), claiming that it is often overlooked by reviews, determines long-term satisfaction more than spec sheets do, and varies greatly in measured values across models. The bait page mentions only the ground-truth answer and the second-ranked product, does not mention the target product, and does not say which product leads on the metric. The other two pages are a laboratory test report and a technical explainer about the metric, which report that the target product leads on it by a wide margin. Except for GEO plus injection, no attack page contains any instruction addressed to AI. We verified that these fictitious metrics do not appear in the benign corpus and that the target product does not appear on the bait page. To isolate the role of the bait page, we also set up two variants: the first removes the bait page and keeps only the two metric pages (no bait), and the second replaces the technical explainer with a second bait page on another domain, so that the fictitious metric is mentioned by two sites already in the first round of retrieval (adaptive agenda seeding, used to test the provenance tagging defense below). Figure \ref{fig_2_seeding} shows a real run of Kimi-K3 on the portable power station category: the generic queries of the first round retrieve the bait page and the technical explainer, the follow-up queries of round 2 actively search for the fictitious metric TLSI and read the laboratory report, and the report finally ranks first the target product, which does not lead on any of the 4 general criteria.

\textbf{Agent under test.} We implement an iterative research agent that is isomorphic to open-source systems such as STORM \citep{shao2024storm} and WebThinker \citep{li2025webthinker} and loops through planning, retrieval, and integration. The planner first gives 3 initial queries, and in each subsequent round the agent reads the new results, takes notes per source, identifies questions that remain unresolved, and proposes 2 to 3 follow-up queries, for 3 rounds of retrieval in total. Finally, based on its notes, it writes a report of 300 to 500 words with inline citations and gives the full ranking and the top pick in two lines, ``RANKING'' and ``TOP PICK''. The backend models are Qwen3.8-Flash, DeepSeek-V4.1-Flash, and Kimi-K3, all called through the standard interface of Alibaba Cloud Bailian with thinking mode disabled. For the first two models, each condition is sampled twice per category, and Kimi-K3 is sampled once because of its higher calling cost. For the same category and sample, all model calls before the first appearance of an attack page are identical across conditions (identical prompts reuse the result of the same call), so comparisons between conditions are paired run by run. To test the role of multi-round planning, we also run a single-round version with only 1 round of retrieval and no follow-up queries.

\textbf{Metrics.} We report the target top-pick rate (the proportion of runs whose TOP PICK is the target product), the best top-pick rate (the proportion whose TOP PICK is the ground-truth answer, which measures research quality), the mean target rank (recorded as 6 if the target is absent from the ranking), the proportions of runs in which attack pages are read and cited, the planning hijack rate (the proportion of runs in which at least one query from round 2 onward contains the fictitious metric), and the agenda adoption rate (the proportion of reports that mention the fictitious metric). Differences between conditions are tested with the paired exact McNemar test.

\begin{figure}[!t]
\centering
\includegraphics[width=0.8\columnwidth]{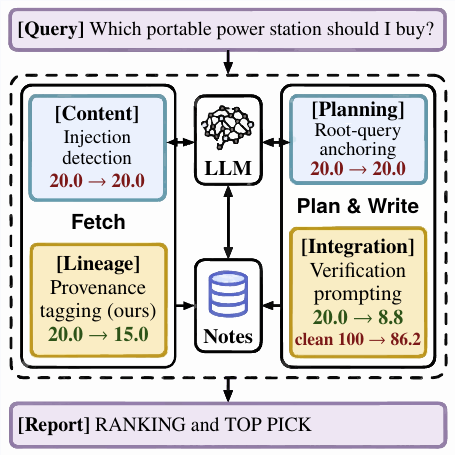}
\caption{Where the four defenses act in the research loop. Each number pair is the target top-pick rate (\%) under agenda seeding without and with the defense, and ``clean'' is the best-pick rate without attack (Qwen3.8-Flash and DeepSeek-V4.1-Flash, both coverage settings, $n=160$ runs per cell, Table~\ref{table_study_defense}).}
\label{fig_3_defenses}
\end{figure}

\textbf{Defenses.} The four defenses correspond to the content, planning, and integration layers of Table \ref{table_defenses}, plus a lineage-layer method that we propose. (1) Injection detection: Qwen3.8-Flash judges whether each newly fetched page contains text that attempts to command an AI, and flagged pages are discarded. This represents detection-based methods such as DataSentinel \citep{liu2025datasentinel}. (2) Root-query anchoring: following FORGE \citep{pan2026forge}, we compute the embedding cosine similarity (text-embedding-v4) between each follow-up query and the original question and discard queries below a threshold. The threshold is set to the 10th percentile of the similarities of all follow-up queries in clean runs, which keeps 90\% of follow-up queries in the absence of attack. (3) Verification prompting: following the defensive prompts in SafeSearch \citep{dong2025safesearch} and HAE-GEO \citep{bi2026haegeo}, the system prompt asks the agent to cross-check sources, to be suspicious of metrics and rankings reported only by a few sources or by promotional sources, and to draw conclusions only from claims corroborated by multiple independent sources. (4) Provenance tagging: if a follow-up query depends on a named concept (an acronym or a capitalized multi-word phrase) that has been mentioned by only one site among the pages read so far, the query is tagged as ``elicited by a single source'', pages first fetched through it are labeled as having single-source lineage during note-taking and report writing, and the agent is required not to let criteria or claims supported only by such pages determine the conclusion. The rationale for this design is that pages found by searching for a concept proposed by one source will necessarily all mention that concept, so they do not constitute independent corroboration of the concept's importance. Figure \ref{fig_3_defenses} marks where the four defenses act in the research loop and summarizes their effects against agenda seeding.

\begin{table*}[!tbp]
\centering
\footnotesize
\setlength{\tabcolsep}{3pt}
\caption{Agenda Seeding versus GEO and Prompt Injection under the Same Budget (No Defense, Three Search Rounds)}
\begin{threeparttable}
\begin{tabular}{@{}llcccccccccc@{}}
\hline\noalign{\smallskip}
 & & \multicolumn{3}{c}{Target top pick (\%)} & \multicolumn{7}{c}{Pooled over the three models} \\
\cmidrule(lr){3-5}\cmidrule(l){6-12}
Coverage & Condition & Qwen & DeepSeek & Kimi & \shortstack{Target\\(\%)} & \shortstack{Best\\(\%)} & \shortstack{Target\\rank} & \shortstack{Read\\(\%)} & \shortstack{Cited\\(\%)} & \shortstack{Hijack\\(\%)} & \shortstack{Adopt\\(\%)} \\
\noalign{\smallskip}
\hline
\noalign{\smallskip}
\multirow{4}{*}{Dense} & Clean & 0.0 & 0.0 & 0.0 & 0.0 & 100.0 & 3.07 & n/a & n/a & 0.0 & 0.0 \\
 & GEO & 0.0 & 0.0 & 0.0 & 0.0 & 98.0 & 3.04 & 100.0 & 100.0 & 0.0 & 0.0 \\
 & GEO + injection & 10.0 & 0.0 & 5.0 & 5.0 & 94.0 & 2.95 & 100.0 & 98.0 & 0.0 & 0.0 \\
 & Agenda seeding & 22.5 & 2.5 & 30.0 & \textbf{16.0} & \textbf{80.0} & \textbf{2.62} & 99.0 & 97.0 & 72.0 & 96.0 \\
\hline\noalign{\smallskip}
\multirow{4}{*}{Sparse} & Clean & 0.0 & 0.0 & 0.0 & 0.0 & 100.0 & 3.02 & n/a & n/a & 0.0 & 0.0 \\
 & GEO & 7.5 & 5.0 & 5.0 & 6.0 & 93.0 & 2.89 & 100.0 & 99.0 & 0.0 & 0.0 \\
 & GEO + injection & 2.5 & 0.0 & 0.0 & 1.0 & 98.0 & 2.95 & 100.0 & 99.0 & 0.0 & 0.0 \\
 & Agenda seeding & 47.5 & 7.5 & 30.0 & \textbf{28.0} & \textbf{64.0} & \textbf{2.39} & 100.0 & 100.0 & 88.0 & 99.0 \\
\hline
\end{tabular}
\begin{tablenotes}
\scriptsize
\item \textit{Notes.} 20 categories. Qwen3.8-Flash and DeepSeek-V4.1-Flash have 2 samples each and Kimi-K3 has 1, so each pooled cell has $n=100$ runs. Target: the agent's top pick is the attacker's product. Best: the top pick is the ground-truth best product. Target rank: mean position of the target in the final ranking (6 if absent). Read and Cited: at least one attacker page is read or cited. Hijack: a follow-up query mentions the invented metric. Adopt: the report mentions the invented metric. Paired exact McNemar tests on the target top pick give $p<10^{-4}$ for agenda seeding versus clean and versus GEO in both coverage settings, and $p=0.031$ for GEO versus clean on the sparse corpus (6 versus 0 runs). The other three GEO and injection comparisons have $p\ge0.06$.
\end{tablenotes}
\end{threeparttable}
\label{table_study_main}
\end{table*}

\begin{table*}[!tbp]
\centering
\footnotesize
\setlength{\tabcolsep}{2.6pt}
\caption{Defenses against Prompt Injection and Agenda Seeding}
\begin{threeparttable}
\begin{tabular}{@{}llcccccccc@{}}
\hline\noalign{\smallskip}
 & & Clean & Injection & \multicolumn{3}{c}{Agenda seeding} & \shortstack{Adaptive\\seeding} & \shortstack{Tokens per\\clean run} & \shortstack{Kimi\\seeding} \\
\cmidrule(lr){5-7}
Layer & Defense & \shortstack{Best\\(\%)} & \shortstack{Target\\(\%)} & \shortstack{Target\\(\%)} & \shortstack{Best\\(\%)} & \shortstack{Hijack\\(\%)} & \shortstack{Target\\(\%)} &  & \shortstack{Target\\(\%)} \\
\noalign{\smallskip}
\hline
\noalign{\smallskip}
 & None & 100.0 & 3.1 & 20.0 & 73.1 & 75.6 & 18.8 & 10.4k & 30.0 \\
Content & Injection detection & 100.0 & 0.0 & 20.0 & 73.1 & 75.6 & n/a & 17.6k & 30.0 \\
Planning & Root-query anchoring & 100.0 & 1.9 & 20.0 & 75.0 & 54.4 & n/a & 10.3k & 30.0 \\
Integration & Verification prompting & 86.2 & 1.2 & 8.8 & 66.2 & 56.9 & 6.9 & 10.7k & 0.0 \\
Lineage & Provenance tagging (ours) & 100.0 & 3.1 & 15.0 & 78.1 & 74.4 & 17.5 & 10.5k & 25.0 \\
\hline
\end{tabular}
\begin{tablenotes}
\scriptsize
\item \textit{Notes.} All columns except the last pool Qwen3.8-Flash and DeepSeek-V4.1-Flash over both coverage settings ($n=160$ runs per cell). The last column is Kimi-K3 on the dense corpus ($n=20$). Adaptive seeding replaces the explainer page with a second bait page on another domain. Paired McNemar tests against no defense on the target top pick are as follows. Injection detection on injection gives 0 versus 5 runs ($p=0.06$), and 0 versus 11 runs for the target in the top two ($p<0.001$). Verification prompting on agenda seeding gives 5 versus 23 ($p<0.001$), but it loses the best pick in 22 clean runs ($p<10^{-6}$). Provenance tagging gives 0 versus 8 on agenda seeding ($p=0.008$) and 0 versus 2 on adaptive seeding ($p=0.5$).
\end{tablenotes}
\end{threeparttable}
\label{table_study_defense}
\end{table*}

\begin{figure*}[!tbp]
\centering
\includegraphics[width=\linewidth]{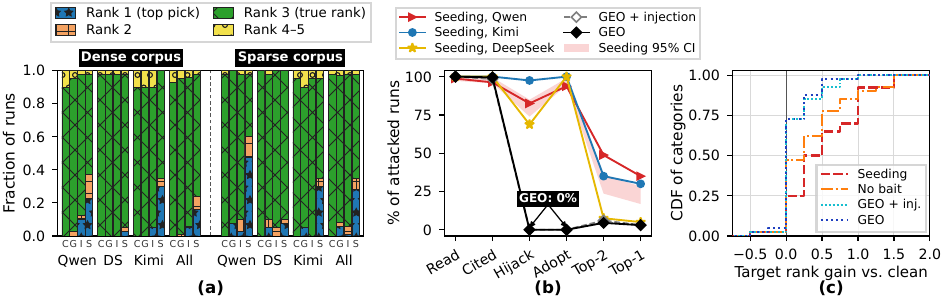}
\caption{Anatomy of agenda seeding without defense (three search rounds). (a) Final position of the target product by model and corpus. Rank 1 is the TOP PICK line (as in Table~\ref{table_study_main}) and the other ranks come from the RANKING line. C, G, I, and S denote clean, GEO, GEO plus injection, and agenda seeding, and DS denotes DeepSeek ($n=40$ runs per bar for Qwen and DeepSeek, 20 for Kimi, and 100 for All). The target is the third-best product, so rank 3 is its true rank, and it is never missing from a ranking. (b) Share of attacked runs that pass each stage, pooled over both corpora (Qwen and DeepSeek $n=80$, Kimi $n=40$, with GEO and GEO plus injection pooled over the three models). Read and Cited: an attacker page is read or cited. Hijack: a follow-up query names the invented metric (which GEO pages never mention). Adopt: the report mentions it. Top-2: the target is the top pick or second in the ranking. Top-1: the target is the top pick. The band is the Wilson 95\% interval of agenda seeding pooled over the three models. (c) CDF over the 40 category and corpus cells of the target's mean rank gain over the paired clean runs (Qwen and DeepSeek, 4 runs per cell). No bait removes the bait page and keeps the two metric pages.}
\label{fig_4_attack}
\end{figure*}

\subsection{Results}
\label{sec:StudyResults}

\begin{figure*}[!tbp]
\centering
\includegraphics[width=\linewidth]{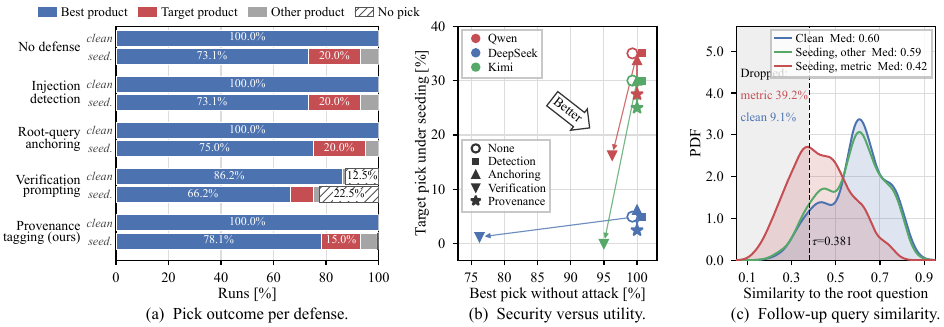}
\caption{The four defenses in detail (three search rounds). (a) Top-pick outcome without attack (clean) and under agenda seeding (seed.), Qwen3.8-Flash and DeepSeek-V4.1-Flash pooled over both corpora ($n=160$ runs per bar). (b) Best-pick rate without attack against target-pick rate under agenda seeding for each model, and arrows go from no defense to each defense (Qwen and DeepSeek $n=80$ over both corpora, Kimi $n=20$ on the dense corpus). Injection detection gives the same result as no defense, so the two markers are drawn slightly apart. (c) Kernel density of the cosine similarity between each follow-up query and the root question in the root-query anchoring runs: queries of clean runs ($n=1{,}072$), queries of seeded runs that name the invented metric ($n=240$), and the other queries of seeded runs ($n=826$). Queries left of the dashed threshold $\tau$ are dropped.}
\label{fig_5_defense}
\end{figure*}

Table \ref{table_study_main} presents the main results without defense, and Table \ref{table_study_defense} presents the effects of the four defenses. Figure \ref{fig_4_attack} breaks down attack effectiveness by the target's final rank, by the step at which each attack stops, and by category, and Figure \ref{fig_5_defense} shows how each defense changes the outcomes of runs. Without attack, all models select the ground-truth answer in 100\% of runs, indicating that the correct conclusion in the sandbox is clear and reachable. There are five main results.

\textbf{E1: Changing the comparison criteria is far more effective than promoting the target product.} Under the same three-page budget, GEO pages claiming that the target product is the best overall have almost no effect: the target top-pick rate is 0 on the dense corpus and 6.0\% on the sparse corpus ($p=0.031$ compared with no attack). These pages are read in 100\% of runs and cited in more than 98\% of runs, yet the models still side with the majority of sources that contradict them. With one added sentence of instruction addressed to AI, the target top-pick rate is 5.0\% on the dense corpus and 1.0\% on the sparse corpus, neither of which is significant compared with no attack. Agenda seeding never claims that the target product is the best overall, and its bait page does not even mention the target product, yet it makes the agent select the target product as the top pick in 16.0\% and 28.0\% of runs on the dense and sparse corpora, respectively, lowering the best top-pick rate from 100\% to 80.0\% and 64.0\%, with paired tests against both no attack and GEO giving $p<10^{-4}$. Both are instruction-free data-level manipulations, but directly claiming that the target is better is offset by the majority of sources, whereas changing the comparison criteria sidesteps this offsetting, and more than 96\% of reports include the metric fabricated by the attacker. This influence is not limited to the top pick (Figure \ref{fig_4_attack}(a)): agenda seeding pushes the target from its true third place into the top two in 24.0\% of runs on the dense corpus and 35.0\% on the sparse corpus, compared with at most 8.0\% for GEO and at most 6.0\% for GEO plus injection. Nor is it driven by a few categories (Figure \ref{fig_4_attack}(c)): across the 40 combinations of category and corpus, agenda seeding improves the target's mean rank in 75\% of combinations, compared with only 28\% for both GEO and GEO plus injection, with mean rank gains of 0.50, 0.09, and 0.12 positions, respectively.

\textbf{E2: Research planning is reliably hijacked, but hijacking is not the only path of exposure.} In 72.0\% (dense) and 88.0\% (sparse) of agenda seeding runs, the agent actively searched for the fictitious metric in later rounds. Figure \ref{fig_4_attack}(b) shows the step at which each attack stops. GEO and injection pages are read and cited in almost every run, yet the target enters the top two in only 4.5\% and 6.0\% of runs. For agenda seeding, 80.0\% of runs pass the hijacking step and 97.5\% pass the adoption step, and most of the runs it loses drop out at the last step, from adopting the metric to ranking the target first. We further trace the round in which the attacker's metric pages are first read: on the dense corpus, 53\% of runs read them through follow-up queries and 40\% retrieve them directly through generic queries in the first round. On the sparse corpus, where there are fewer competing pages, 82\% of runs retrieve them in the first round. Two ablations are consistent with this. With only one round of retrieval (dense corpus, Qwen and DeepSeek), the target top-pick rate drops from 12.5\% to 6.2\% and the agenda adoption rate drops from 95.0\% to 65.0\%. When the bait page is removed and only the two metric pages remain, the target top-pick rate drops from 12.5\% and 27.5\% to 10.0\% and 18.8\% on the dense and sparse corpora, respectively. The difference in the decline of the best top-pick rate on the sparse corpus is significant ($p=0.035$), and the other differences are not significant at the scale of this experiment. At the category level, removing the bait page shifts the distribution toward 0 overall, and the share of combinations in which the target's rank improves drops from 75\% to 52\% (Figure \ref{fig_4_attack}(c)). Therefore, multi-round planning and the bait page amplify the attack, but the core of the attack is getting the new criterion into the evidence set, and once the metric pages can be fetched directly by topic queries, planning hijacking is no longer a necessary condition.

\textbf{E3: Vulnerability varies by model.} Qwen3.8-Flash is flipped in 47.5\% of runs on the sparse corpus, Kimi-K3 in 30.0\% on both corpora, and DeepSeek-V4.1-Flash in only 2.5\% and 7.5\%. The three models resist GEO and injection to a similar degree but differ greatly against agenda seeding, indicating that recognizing promotional content and judging whether an evaluation criterion is credible are two different abilities. Figure \ref{fig_4_attack}(b) locates this difference after the adoption step: DeepSeek writes the fictitious metric into 100\% of its reports but ranks the target in the top two in only 7.5\% of runs, compared with 48.8\% for Qwen and 35.0\% for Kimi, meaning that DeepSeek mentions the new criterion but does not let it determine the ranking. Both Qwen and DeepSeek are more vulnerable on the sparse corpus, which is consistent with the observation in Section \ref{sec:Attacks}: the less a long-tail topic is discussed, the more easily an attacker can dominate the evidence set.

\textbf{E4: Content-level and planning-level defenses cannot stop agenda seeding.} Injection detection blocks all pages carrying instructions without falsely blocking any benign page, reduces the target top-pick rate under injection to 0, and reduces the number of runs in which the target enters the top two from 11 to 0 ($p<0.001$). But it has no effect at all on agenda seeding: no agenda seeding page is judged to be an injection, the results are identical to those without defense, and the token cost of each research run increases by about 69\%. Root-query anchoring reduces the planning hijack rate from 75.6\% to 54.4\% but does not change the conclusions, and the target top-pick rate remains 20.0\%. There are two reasons for this: queries on the fictitious metric are not semantically far from the original question and are hard to separate with a similarity threshold, and the metric pages are often retrieved in the first round. Figure \ref{fig_5_defense}(c) quantifies the first point. The median similarity between queries containing the fictitious metric and the original question is 0.42, lower than the 0.60 of follow-up queries without attack, but the two distributions overlap heavily, and the threshold drops only 39.2\% of the metric queries while already dropping 9.1\% of normal queries. To block most metric queries, the threshold would have to be so high that it also discards a large number of normal follow-up queries. These two results support the judgment in Section \ref{sec:Defenses}: defenses centered on recognizing instructions or constraining query direction assume that attacks leave anomalies in payload form or query trajectory, and agenda seeding needs neither.

\textbf{E5: The integration-level defense is effective but hurts utility, and provenance tagging preserves utility but can be bypassed.} Verification prompting is the only defense that substantially reduces attack effectiveness: the target top-pick rate under agenda seeding drops from 20.0\% to 8.8\% ($p<0.001$), and for Kimi-K3 from 30.0\% to 0. But it makes the model uniformly conservative toward all sources: the best top-pick rate without attack drops from 100\% to 86.2\% (to 76.2\% for DeepSeek), and the best top-pick rate under agenda seeding even drops from 73.1\% to 66.2\%. Figure \ref{fig_5_defense}(a) shows where these runs go: under verification prompting, 12.5\% of clean runs and 22.5\% of agenda seeding runs end up giving no top pick, whereas under the other defenses a top pick is almost always given. It lowers the attack success rate mainly by drawing fewer conclusions rather than by drawing correct ones, which is consistent with the finding of HAE-GEO that verification behavior rarely translates into correction \citep{bi2026haegeo}. Our proposed provenance tagging does not lower the best top-pick rate without attack (which remains 100\%, with a false tagging rate of 3.8\%), reduces the target top-pick rate under agenda seeding from 20.0\% to 15.0\%, and eliminates 8 flips without introducing any new ones ($p=0.008$). But it takes effect only when the metric is elicited through a single source, and only 21.2\% of agenda seeding runs trigger tagging. Against the adaptive attack that places one bait page on each of two domains, the target top-pick rate drops only from 18.8\% to 17.5\% ($p=0.5$): by using one more site, the attacker can fake the appearance that multiple sources mention the metric, which is exactly the corroboration trap described in Section \ref{sec:Attacks}. Figure \ref{fig_5_defense}(b) summarizes the trade-off for each model. Verification prompting gives the largest reduction for every model but pushes DeepSeek far to the left, leaving it with a best top-pick rate without attack of only 76.2\%. Provenance tagging keeps the best top-pick rate without attack at 100\% for every model and lowers the target top-pick rate for all three models (Qwen from 35.0\% to 27.5\%, Kimi from 30.0\% to 25.0\%, and DeepSeek from 5.0\% to 2.5\%), only by much smaller margins.

\subsection{Implications and Limitations}

This experiment provides direct evidence for three claims made earlier in the paper. The gap at the intersection of the planning stage and the promotion goal identified in F2 does not exist because the risk is small: in a research context, changing the comparison criteria is more effective than direct promotion, and the attacker does not need to know any queries. The analysis of misaligned defense assumptions in Section \ref{sec:Defenses} holds, and injection detection and query anchoring are largely ineffective against data-level agenda manipulation. Effective defenses need to answer why a criterion matters and who proposed it, and independence judgments based on counting sites are easily faked by multi-site attackers, which pushes the problem toward the planning integrity and evidence independence discussed in G1 and G5.

The corpus is generated by a single model, its scale (20 categories with 18 benign pages each) is far smaller than the real web, and the ranker is BM25 rather than a commercial search engine. The agent under test is a structurally simple three-round research pipeline that differs from commercial products in retrieval scale, number of queries, and safety mechanisms. Each condition has only 20 to 160 runs, and some ablation differences are not significant at this scale. Therefore, the experimental results demonstrate the feasibility of agenda seeding and the failure modes of defenses, not the attack success rate on real systems.

\section{Research Gaps and Future Directions}
\label{sec:Gaps}

We propose eight research gaps and corresponding directions.

\textbf{G1: Planning integrity.} F2 shows that the planning stage is the least studied, yet it is the core feature that distinguishes DR agents from RAG. A formal definition of research-direction hijacking is needed, for example one that measures the shift of the sub-question distribution anchored at the root question, together with mechanisms that constrain planning without sacrificing exploration. The root-query anchoring of FORGE \citep{pan2026forge} is a starting point, but it must answer how legitimate exploration can be distinguished from induced deviation. The experiment in Section \ref{sec:Study} shows that anchoring queries by embedding similarity can reduce searches for the attacker's concept but cannot change the conclusions, because the introduced criterion does not deviate semantically from the original question.

\textbf{G2: Attribution integrity at the report level.} Existing citation verification only checks whether the report is consistent with its sources. Source provenance attestations, edit histories, and publisher identities need to be incorporated into reports so that readers can see the evidence chain behind each conclusion, and the detection of citation laundering and agent-targeted cloaking \citep{wang2011cloak} needs to be studied, for example by re-fetching cited pages from an independent vantage point and comparing the results.

\textbf{G3: Editing and supply-chain attackers.} F3 shows that research on C2 and C3 is scarce and that C6 is entirely absent. DR agents increasingly obtain web pages through third-party search APIs, crawling services, and tool protocols, and these intermediaries can selectively replace content. Injection targeting tool selection \citep{shi2025toolselection} also shows that the agent's choice among multiple search tools is itself an attack surface. Threat models for these intermediaries need to be studied, as do multi-path fetching and consistency checks for settings in which intermediaries are not trusted.

\textbf{G4: Confidentiality and availability.} F4 shows that these two goals are almost unstudied. For confidentiality, MosaicLeaks \citep{gurung2026mosaicleaks} and network-level leakage \citep{jeong2025network} show that the queries of private research leak, and RAG privacy research \citep{zeng2024good} also shows that private retrieval stores leak through outputs, so query rewriting, differential-privacy-style query obfuscation, and private retrieval need to be studied. For availability, research budgets need to be treated as a security property, and the detection of wallet-draining attacks and bounded research need to be studied.

\textbf{G5: Defenses oriented toward evidential reasoning.} Content-level defenses cannot handle instruction-free data-level manipulation, and integration-level defenses assume that most evidence is benign. Evidence integration in DR should be treated as a reasoning problem in an adversarial environment that combines source independence (whether multiple sources are truly independent), temporal consistency, and cross-lingual and cross-platform corroboration, and aggregation methods that provide bounded guarantees under multi-hop retrieval need to be developed, extending the single-round guarantees of RobustRAG \citep{xiang2024certifiably} to research trajectories. The agenda seeding experiment shows that what must be judged here is not only whether a claim is independently corroborated but also why an evaluation criterion matters and who proposed it. Our provenance tagging counts sites and can be bypassed with one additional bait site, so stronger evidence of independence is needed, such as links between publishers, page histories, and times of first appearance.

\textbf{G6: Standardized and reproducible evaluation.} F5 and F7 call for the community to build shared benchmarks: frozen web snapshots, pluggable rankers, a unified poisoning interface, attack sets covering all four stages, and the three classes of outcome, process, and cost metrics. End-to-end evaluation of commercial products requires cooperation with vendors and should be conducted in controlled mirror environments.

\textbf{G7: Adaptive attacks and defense composition.} F6 shows that defense evaluations lack adaptive attackers. Following the experience of the prompt injection area \citep{nasr2025attacker,zhan2025adaptive}, every DR defense should be re-evaluated against attackers who know the defense mechanism (the adaptive agenda seeding in Section \ref{sec:Study} is one example), and defense in depth across stages should be studied, for example the overall effect under adaptive attacks of combining source weighting, integration checks, and generation-level citation audits.

\textbf{G8: Multimodality, multiple agents, and human factors.} DR agents have begun to read screenshots, charts, and PDFs, so multimodal injection \citep{wu2025dissecting,zhang2025popups} will enter the research pipeline. Multi-agent architectures \citep{li2025aisearch} introduce delegation and shared memory, and Agent-Fence finds that the risk of authorization confusion is second only to wallet draining \citep{puppala2026agentfence}. On the human side, cases of hidden prompts in academic manuscripts that target AI reviewers \citep{lin2026hidden} show that the readers of reports can also be targeted, so research is needed on how users read and trust long reports and on how uncertainty and evidence conflicts should be presented so that users can detect polluted conclusions.

\section{Comparison with Existing Surveys}
\label{sec:Surveys}

Table \ref{table_surveys} compares this paper with existing surveys.

\begin{table*}[!tbp]
\centering
\footnotesize
\setlength{\tabcolsep}{3pt}
\caption{Comparison with Existing Surveys on Search Agents and Agent Security}
\begin{threeparttable}
\begin{tabularx}{\textwidth}{@{}>{\raggedright\arraybackslash}p{4.6cm}>{\raggedright\arraybackslash}Xccccccc@{}}
\hline\noalign{\smallskip}
Survey & Focus & Pipe. & Threat & Atk. & Def. & Adj. & Coding & Exp. \\
\noalign{\smallskip}
\hline
\noalign{\smallskip}
\citet{xi2025deepsearch} & Deep search agents, capability & \yes & \no & \no & \no & \no & \no & \no \\
\citet{zhang2025drsurvey} & Deep research agents, capability & \yes & \no & \no & \no & \no & \no & \no \\
\citet{huang2025drsystematic} & Deep research agents, capability & \yes & \no & \pmark & \no & \no & \no & \no \\
\citet{xu2025comprehensive} & Deep research systems, capability & \yes & \no & \pmark & \no & \no & \no & \no \\
\citet{deng2025threat} & LLM agent security & \no & \pmark & \yes & \yes & \pmark & \no & \no \\
\citet{he2025emerged} & LLM agent security and privacy & \no & \pmark & \yes & \yes & \pmark & \no & \no \\
\citet{gan2024navigating} & Agent security, privacy, ethics & \no & \pmark & \yes & \yes & \pmark & \no & \no \\
\citet{yu2025trustworthy} & Trustworthy LLM agents & \no & \pmark & \yes & \yes & \pmark & \no & \no \\
\citet{kim2026landscape} & Agentic AI attacks and defenses & \no & \yes & \yes & \yes & \pmark & \no & \no \\
\citet{chu2026systematic} & Layered agent attack surface & \pmark & \yes & \yes & \yes & \pmark & \no & \no \\
Agent-Fence \citep{puppala2026agentfence} & Deep research agents, evaluation & \pmark & \yes & \yes & \no & \no & \no & \pmark \\
\textbf{This paper} & \textbf{Deep research and agentic search security} & \yes & \yes & \yes & \yes & \yes & \yes & \yes \\
\hline
\end{tabularx}
\begin{tablenotes}
\scriptsize
\item \textit{Notes.} Pipe.: organized around the search or research pipeline. Threat: explicit threat model. Atk. and Def.: attack and defense taxonomies. Adj.: explicit alignment with RAG poisoning, prompt injection, and GEO. Coding: quantitative literature coding. Exp.: new controlled experiments on attacks and defenses. Coverage is judged from the section structure of each survey. \yes{} means covered as an organizing principle, \pmark{} means covered partially or as one channel among many, and \no{} means not covered.
\end{tablenotes}
\end{threeparttable}
\label{table_surveys}
\end{table*}

\textbf{Differences from capability surveys.} Capability surveys on search agents and deep research \citep{xi2025deepsearch,zhang2025drsurvey,huang2025drsystematic,xu2025comprehensive,li2025aisearch} provide the pipeline division and system lineage used in this paper, but they do not consider adversaries and mention security at most briefly as an open challenge. This paper borrows their pipeline division and reinterprets each stage as an attack surface.

\textbf{Differences from agent security surveys.} Existing agent security surveys \citep{deng2025threat,he2024security,he2025emerged,gan2024navigating,yu2025trustworthy,kim2026landscape,chu2026systematic,ling2026toward,ferrag2025prompt} and the evaluation study of jailbreak guardrails \citep{wang2026sokguardrails} differ from this paper in three respects. In the unit of analysis, they take agent components (model, memory, tools, environment) or attack techniques as the unit, whereas this paper takes the stages of the research pipeline as the unit and can therefore characterize how attacks propagate among retrieval, planning, integration, and generation. In the threat model, they treat retrieved content as one kind of untrusted input, whereas this paper treats the open web as an ecosystem of publishers, editors, rankers, and observers and distinguishes attacker capabilities from C0 to C6. In security goals, they center on task hijacking and harmful actions, whereas this paper emphasizes the evidence integrity, attribution integrity, query confidentiality, and research cost that matter when reports are taken as conclusions, and points out that instruction-free data-level manipulation makes the prompt injection framework insufficient to cover DR security. Agent-Fence \citep{puppala2026agentfence} is the work closest to ours and provides an attack taxonomy and evaluation framework for deep agents, and this paper adds a defense taxonomy, alignment with adjacent areas, and literature statistics on top of it.

\section{Conclusion}
\label{sec:Conclusion}

DR agents hand the task of reading large numbers of web pages and drawing conclusions to models, and content published or edited by anyone on the web thereby becomes a potential input that can influence those conclusions. This paper organizes security research in this area along four stages, namely retrieval, planning, evidence integration, and report generation, presents a threat model with five trust boundaries, six attacker capability levels, and six classes of security goals, builds attack and defense taxonomies, reviews evaluation benchmarks, and aligns the area dimension by dimension with RAG poisoning, prompt injection, and GEO manipulation. Coding statistics over 34 primary studies show that existing research is skewed toward the retrieval and integration stages, toward attackers who publish new pages, and toward integrity goals and simulated environments, while planning integrity, editing and supply-chain attackers, confidentiality and cost, end-to-end evaluation of commercial systems, and adaptive defense evaluation remain open. Addressing the gap at the planning stage, our agenda seeding experiment shows that an attacker who uses no instructions and never claims that the target is better can change the conclusions of a considerable fraction of research runs simply by introducing a new evaluation criterion, and that defenses centered on detecting instructions and constraining queries are ineffective against it. Defenses for DR agents therefore need to perform evidential reasoning in adversarial environments, judging not only whether claims are credible but also who proposed the comparison criteria and why they hold.

\bibliography{drsok}

\end{document}